\documentclass[a4paper,12pt]{article}
\pdfoutput=1 % if your are submitting a pdflatex (i.e. if you have
\usepackage{jheppub} % for details on the use of the package, please
\usepackage{tabularx,subcaption,fancyhdr,mathtools,physics}
\usepackage[makeroom]{cancel}
\usepackage{bm}
\usepackage{csquotes}
\usepackage[nottoc,notlot,notlof]{tocbibind}
\usepackage{float}
\usepackage{relsize}
\usepackage{hyperref}
\usepackage{color,latexsym, array,multirow,url,verbatim, enumerate,cancel}
\DeclareUnicodeCharacter{2212}{-}
\hypersetup{colorlinks=true,linkcolor=blue,citecolor=magenta,filecolor=magenta,
	urlcolor=cyan}
\def \lspone{\widetilde\chi_1^0}
\def \mlspone{m_{\lspone}}
\def \lsptwo{\widetilde\chi_2^0}

\def\chonep{\widetilde{\chi}_1^{+}}
\def\chonem{\widetilde{\chi}_1^{-}}
\def\chonepm{\widetilde{\chi}_1^{\pm}}
\def\chonemp{\widetilde{\chi}_1^{\mp}}

\def \met{\rm E{\!\!\!/}_T}

\newcommand{\beq}{\begin{equation}}
\newcommand{\eeq}{\end{equation}}
\def\bea{\begin{eqnarray}}
\def\eea{\end{eqnarray}}

\title{\boldmath Status of R-parity violating SUSY}

\author[a]{Arghya Choudhury,} 
\author[a]{Arpita Mondal,}
\author[b]{and Subhadeep Mondal}

\affiliation[a]{Department of Physics, Indian Institute of Technology Patna, Bihar - 801106, India}
\affiliation[b]{Department of Physics, SEAS Bennett University, Greater Noida, Uttar Pradesh  -201310, India}
\emailAdd{arghya@iitp.ac.in}
\emailAdd{arpita\_1921ph15@iitp.ac.in}
\emailAdd{subhadeep.mondal@bennett.edu.in}

\abstract{In this article, we discuss various phenomenological implications of possible R-parity violating (RPV) supersymmetric scenarios. In this context, the implications of both bilinear and trilinear RPV terms are reviewed from the viewpoint of neutrino physics, anomalous muon magnetic moment, different flavor observables, and collider physics. Apart from discussing the distinctive phenomenological implications of the RPV scenarios, we also survey the updated results from different studies to highlight the present status of the RPV couplings.}

\begin{document}
\maketitle

\flushbottom

\setlength{\parskip}{1.0em}

%%%%%%%%%%%%%%%%%%%%%%%%%
\section{Introduction}
\label{sec:intro}
%%%%%%%%%%%%%%%%%%%%%%%%%
In the aftermath of the Higgs boson discovery~\cite{ATLAS:2012yve, CMS:2012qbp}, supersymmetry (SUSY) \cite{Martin:1997ns,drees2004theory,baer2006weak, Susskind:1982mw} rose above all the other proposed beyond the standard model (BSM) scenarios as the favored model. Among various possible SUSY scenarios, the minimal supersymmetric standard model (MSSM) has been studied the most extensively \cite{KumarBarman:2020ylm, Barman:2022jdg,He:2023lgi,Chakraborti:2015mra,Chakraborti:2014gea,Chakraborti:2017dpu,Chowdhury:2016qnz,Bhattacharyya:2011se,Choudhury:2012tc,Choudhury:2013jpa,Baer:2021aax,Athron:2021iuf,Endo:2021zal,Chakraborti:2021bmv,Choudhury:2017acn,Chakraborti:2021dli,Kowalska:2015zja,Choudhury:2016lku, Dutta:2015exw,Dutta:2017jpe,Barman:2016kgt,cms_susy,atlas_susy}. In this minimal SUSY extension, a discrete symmetry, known as the R-parity ($R_p = (-1)^{3B-L+2S}$, where, $B$, $L$ and $S$ represent the baryon number, lepton number and the spin of the particle) \cite{Weinberg:1981wj,Sakai:1981pk,Dimopoulos:1981dw,Farrar:1978xj}, is assumed to be an exact symmetry of the theory. One of the primary motivations for this is to ensure the stability of the lightest SUSY particle (LSP), which can then be a possible dark matter (DM) candidate \cite{KumarBarman:2020ylm, Barman:2022jdg,He:2023lgi, Chakraborti:2015mra, Chakraborti:2014gea, Chakraborti:2017dpu, Chowdhury:2016qnz,Bhattacharyya:2011se,Choudhury:2012tc,Choudhury:2013jpa}. Under R-parity, all the SM particles are even, and all the SUSY particles are odd. Hence, within the RPC MSSM framework, the SUSY particles can only be produced in pairs and cannot decay exclusively to the SM particles.
All RPC SUSY decays, therefore, end up with the LSP. This also results in large transverse missing energy being associated with characteristic R-parity conserving (RPC) MSSM signatures at the collider experiments \cite{ATLAS:2022hbt,ATLAS:2022ckd,ATLAS:2022zwa,CMS:2021few,CMS:2022vpy,CMS:2022sfi}. However, no such signal has so far been observed at the large hadron collider (LHC), which puts the minimal version under strain. This prompts us to explore other SUSY scenarios which may be more viable options. One of the simplest extensions of the MSSM is to allow R-parity violating terms in the lagrangian, which gives rise to four additional gauge invariant terms in the superpotential.

The R-parity violating superpotential~\cite{Barbier:2004ez} can be written as:
\begin{equation}
\label{eq:rpv_potential}
W_{\cancel{R}_p} = \epsilon_i \hat{L}_i \hat{H}_u + \frac{1}{2}\lambda_{ijk}\hat{L}_i\hat{L}_j\hat{E}_k^c + \lambda^{\prime}_{ijk}\hat{L}_i\hat{Q}_j\hat{D}_k^c + \frac{1}{2}\lambda^{\prime\prime}_{ijk}\hat{U}_i^c\hat{U}_j^c\hat{D}_k^c 
\end{equation}
The first bilinear term with $\epsilon_i$ and the next two trilinear terms containing $ \lambda$ and $\lambda^{\prime}$ couplings in the Eq.~\ref{eq:rpv_potential} violate lepton number by one unit and the last trilinear term with $\lambda^{\prime\prime}$ coupling violates the baryon number by one unit. Here $H_u$ represents the supermultiplet of up-type Higgs. $\hat{Q}_j$, $\hat{U}_j$ ($\hat{D}_k$) represent left-handed doublet and right-handed singlet up-type (down-type) quark supermultiplet, respectively. Similarly, $\hat{L}_i$ and $\hat{E}_k$ characterize the left-handed and right-handed lepton supermultiplet respectively.
Allowing R-parity violation gives rise to an array of phenomenological consequences. 

One of the major motivations for looking for BSM physics is the observation of light neutrino masses and mixing. A class of neutrino mass models suggests that one must have a local $(B-L)$ violating term in the lagrangian in order to generate neutrino masses \cite{Hayashi:1984rd,Mohapatra:1986aw,Barger:2008wn,FileviezPerez:2008sx}. Supersymmetrization of these theories leads to RPV scenarios. One can use the bilinear term to generate one non-zero neutrino mass at tree level \cite{Banks:1995by,Grossman:1998py,Nardi:1996iy,Rakshit:2004rj}. The other neutrino masses are generated at one loop level through the bilinear as well as the lepton number violating trilinear terms ($\lambda$ and $\lambda^{\prime}$) \cite{Davidson:2000uc,Davidson:2000ne,Rakshit:2004rj,Hall:1983id, Babu:1989px, Barbier:2004ez,Dreiner:2022zsc}. Given the solar and atmospheric neutrino mass scales, this puts stringent constraints on RPV parameters \cite{Hirsch:2000ef,Abada:2001zh, Diaz:2004fu,Hempfling:1995wj,Gozdz:2008zz,Choudhury:2023lbp}. A consequence of having non-zero RPV couplings is that the LSP can now decay into SM particles. However, the couplings can be sufficiently small to ensure that the LSP has a sufficiently long lifetime and thereby still remains a viable DM candidate \cite{Restrepo:2011rj,Buchmuller:2007ui,Ibarra:2007jz,Arnold:2013qja,Cottin:2014cca}. 

RPV couplings affect various high as well as low-energy processes. The presence of the trilinear couplings gives rise to additional contributions to charged current and neutral current interactions which impacts precision observables such as the Fermi coupling, $W$ and $Z$ boson masses, their decay widths, $\rho$ parameter, Weinberg angle, to name a few \cite{Dimopoulos:1988jw,Erler:1998ig,Amaldi:1987fu,Costa:1987qp}. The $\lambda$ couplings give rise to both two-body and three-body lepton number violating processes \cite{Amaldi:1987fu,Costa:1987qp,Barger:1989rk}. No such processes have been observed to date, which puts restrictions on these couplings \cite{Amaldi:1987fu,Costa:1987qp,Barger:1989rk}. The other lepton number violating trilinear coupling $\lambda^{\prime}$ contributes to neutrinoless double beta decay, which violated lepton number by two units \cite{Bednyakov:1996ps}. These $\lambda^{\prime}$ couplings also affect the CKM matrix elements and precisely measured observables like $R_D$ and $R_D^*$ \cite{Altarelli:1997qu,Grossman:1995gt,Erler:1996ww,Dreiner:2006gu}. Constraints can be drawn on the $\lambda$ and $\lambda^{\prime}$ couplings from studies of forward-backward asymmetries in scatterings such as $e^+ e^-\to f\bar f$, where, $f\equiv l, q$ \cite{Barbier:2004ez}. 
Measurement of muon magnetic moment provides another indication of possible BSM physics \cite{Muong-2:2006rrc,Muong-2:2021vma,Muong-2:2021ojo,Muong-2:2023cdq}. Recent measurements at the Fermi lab puts the global average of muon (g-2) beyond $5\sigma$ of the SM prediction \cite{Muong-2:2023cdq}. RPV MSSM can provide a substantial contribution to the magnetic moment calculation over that of the RPC contribution through both bilinear and trilinear terms \cite{Martin:2001st,Moroi:1995yh,Hundi:2011si,Kim:2001se,Chakraborty:2015bsk}. Existing matter anti-matter asymmetry in nature can be explained through additional sources of CP-violation over that of the SM. The possible connection between RPV and the CP-violation has been studied extensively \cite{Domingo:2018qfg}. Studies of $K\bar{K}$ oscillation, hadronic decay of $B$ mesons, electric dipole moment of fermions, etc. can put additional constraints on the RPV couplings \cite{Domingo:2018qfg}. 

Both ATLAS and CMS collaborations have explored various RPV models with simplified scenarios and in the process have put constraints on the SUSY particle masses \cite{ATLAS:2021fbt,ATLAS:2020wgq,ATLAS:2019fag,ATLAS:2018umm,CMS:2021knz,CMS:2018skt,CMS:2017szl,CMS:2016vuw,CMS:2016zgb,CMS:2013pkf,ATLAS:2015gky,ATLAS:2015rul} which are widely different from that of the RPC constraints. The major difference in the signals of RPV and RPC lies in the fact that in the latter case, the LSP cannot decay and hence is associated with a large missing energy, which is lacking in the former. However, the large missing energy is traded with a larger multiplicity of the leptons and jets in the final state, which is equally, if not more, effective in probing the SUSY particles. Choosing different configurations of RPV couplings and different types of the LSP gives rise to a large number of possible experimental signatures \cite{Dreiner:2023bvs,Barman:2020azo,Mitsou:2015kpa, Mitsou:2015eka, Bardhan:2016gui,Bhattacherjee:2013tha,Bhattacherjee:2013gr,Dercks:2018eua,Cohen:2019cge,Choudhury:2023eje, Choudhury:2023yfg}. These studies do not aim to constrain the RPV couplings. They simply assume that the couplings are large enough for the SUSY particles to decay promptly. However, if the couplings are constrained from some other observations to be small enough, prompt decays may not take place. This gives rise to long-lived particles, and for them, the limits change significantly \cite{CMS:2016vuw,ATLAS:2023oti, CMS:2020iwv, Bhattacherjee:2023kxw}. 

In this article, we aim to briefly discuss all these phenomenological implications of the RPV SUSY scenario and survey their present status. In Section~\ref{sec:model}, we discuss the model framework. In Section~\ref{sec:neutrino}, we present how neutrino masses are generated within this model framework. In Sec.~\ref{sec:mug2} and \ref{sec:flavor}, we summarize how low energy observables affect the choice of RPV couplings. In Section~\ref{sec:collider}, collider implications of various RPV models and existing limits on the SUSY masses are discussed. The future prospects of electroweakino production at the HL-LHC and HE-LHC for different
RPV couplings are presented in Sec.~\ref{sec:electroweakino_pheno}. Finally, we conclude the paper with a summary in Sec.~\ref{sec:concl}.  
\begin{figure}[h]%
\centering
\includegraphics[width=6cm,height=3.5cm]{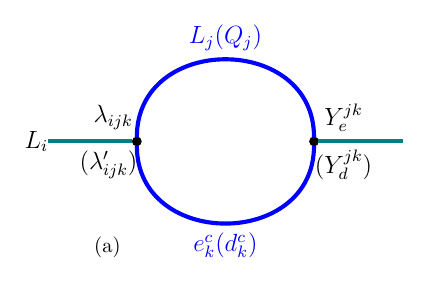} 
\hspace{0.7cm}
\includegraphics[width=6cm,height=3.5cm]{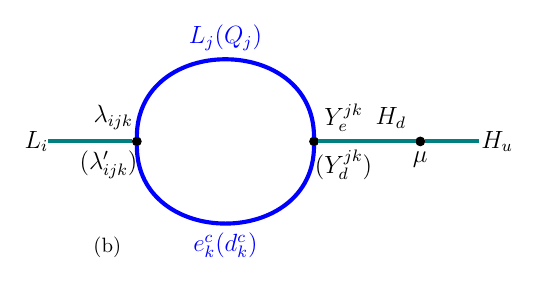}
\caption{Generation of the bilinear RPV term from non-zero trilinear $\lambda$ or $\lambda^{\prime}$ couplings.}
\label{fig:bi_tri}
\end{figure}
%%%%%%%%%%%%%%%%%%%%%%%%%%%%%%%%
\section{Model description}
\label{sec:model}
%%%%%%%%%%%%%%%%%%%%%%%%%%%%%%%%
Eq.~\ref{eq:rpv_potential} shows the additional terms that can be added to the MSSM superpotential. The bilinear extension is the simplest one. One can, in principle, get rid of this term through a redefinition of the Higgs and lepton superfields \cite{deCampos:1995ten}. However, a rotation like $\hat{H}_d\to \epsilon\hat{H}_d+\epsilon_i\hat{L}_i$ can give rise to RPV scalar mass terms starting from a RPC SUSY breaking term and as a result one has bilinear RPV terms in the scalar potential \cite{deCampos:1995ten}. Note that these bilinear terms can also show up at some other energy scale even if they are rotated away to start with \cite{Barger:1995qe}. The trilinear couplings can also give rise to these bilinear terms \cite{deCarlos:1996ecd} as shown in Fig.~\ref{fig:bi_tri}. 
On the other hand, trilinear interactions similar to the $\lambda$ and $\lambda^{\prime}$ couplings can also be generated starting from non-zero bilinear couplings \cite{Roy:1996bua}. 

\begin{figure}[h]%
\centering
\includegraphics[width=0.5\textwidth]{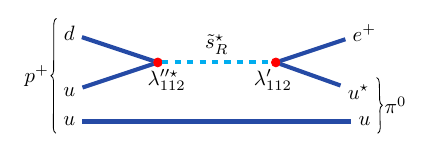}
\caption{Proton decay into $e^+\pi$ final state in presense of $\lambda^{\prime}$ and $\lambda^{\prime\prime}$ couplings through squark mediation.}
\label{fig:proton_decay}
\end{figure}

In the standard model as well as in RPC MSSM, the lepton number and baryon numbers are conserved. These conservations are not a consequence of any fundamental symmetries of nature. However, in the presence of both non-zero lepton number and baryon number violating couplings, we obtain a robust phenomenological constraint from proton decay. A proton can decay via $p\to l^+\pi$, where $l=e^+,~\mu^+$, in presence of $\lambda^{\prime}$ and $\lambda^{\prime\prime}$ couplings as shown in Fig.~\ref{fig:proton_decay}.
The known proton lifetime is $> 10^{32}$ years. This puts a constraint on the couplings and squark masses, $\frac{|\lambda^{\prime 11i}\lambda^{\prime\prime 11i}|}{m^2_{\tilde d_i}}~({\rm GeV})^2 < 2\times 10^{-31}$. There are other processes that put robust constraints on the RPV couplings \cite{Zwirner:1983dgv,Barger:1989rk,Godbole:1992fb,Bhattacharyya:1995pq,Bhattacharyya:1996nj,Dreiner:1997uz,Allanach:2003eb}. 

\begin{figure}[h]%
\centering
\includegraphics[width=0.32\textwidth]{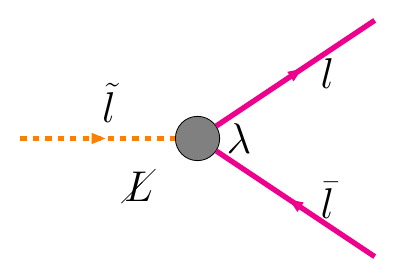}
%\hspace{0.3cm}
\includegraphics[width=0.32\textwidth]{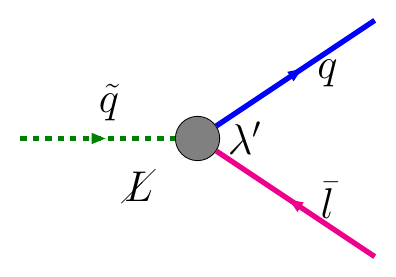}
%\hspace{0.cm}
\includegraphics[width=0.32\textwidth]{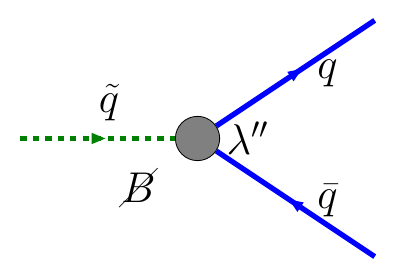}
\caption{Here the tree diagrams of trilinear RPV couplings corresponding to $\lambda$ (Left), (Middle) $\lambda^{\prime}$ (Middle), and $\lambda^{\prime\prime}$ (Right) are shown. The arrows show the flow of baryon and lepton numbers. Here $q$ ($\tilde{q}$) corresponds to quark (squark) and similarly $l$ ($\tilde{l}$) represents lepton(slepton). $\cancel{L}$ and $\cancel{B}$ corresponds to the lepton and baryon number violation respectively.}
\label{fig:couplings}
\end{figure}

The way the RPV terms are written in Eq.~\ref{eq:rpv_potential} summation over $i,~j,~k=1,2,3$ indices are implied. Consequently, there are three bilinear coupling parameters. The $\lambda$ term can be rewritten with explicit SU(2)$_L$ indices as $\frac{1}{2}\lambda_{ijk}\epsilon_{ab}\hat{L}^a_i\hat{L}^b_j\hat{E}_k^c$, where, $a,~b=1,2$ and $\epsilon_{ab}$ is an anti-symmetric tensor. As a result, one can write, 
\bea
\lambda_{ijk}\hat{L}_i\hat{L}_j\hat{E}_k^c = \lambda_{ijk}\epsilon_{ab}\hat{L}^a_i\hat{L}^b_j\hat{E}_k^c = -\lambda_{jik}\epsilon_{ba}\hat{L}^a_j\hat{L}^b_i\hat{E}_k^c = -\lambda_{jik}\hat{L}_i\hat{L}_j\hat{E}_k^.
\eea
This makes the $\lambda_{ijk}$ couplings anti-symmetric with respect to the first two indices, $\lambda_{ijk} = -\lambda_{jik}$. So there are 9 independent $\lambda_{ijk}$ couplings. $\lambda_{ijk}^{\prime}$ has no such property and it has 27 independent couplings. Similar to the $\lambda_{ijk}$, gauge invariance ensures that $\lambda^{\prime\prime}_{ijk}$ couplings are anti-symmetric in the last two indices, $\lambda^{\prime\prime}_{ijk} = -\lambda^{\prime\prime}_{ikj}$. As a result, we have 9 independent $\lambda^{\prime\prime}_{ijk}$ couplings. Fig.~\ref{fig:couplings} depicts the three trilinear coupling vertices as they appear in Eq.~\ref{eq:rpv_potential}.
%%%%%%%%%%%%%%%%%%%%%%%%%%%%%%%%%%%%%%%%%%%%%%%%%%%%%%%%%%%%%%%%%%%%%%%%%%%%%%%%%%%%%%%%%%%%%%%%%%%%%%%%
\section{Neutrino Physics}
\label{sec:neutrino}
%%%%%%%%%%%%%%%%%%%%%%%%%%%%%%%%%%%%%%
This section reviews the prospect of light neutrino mass generation within the framework of R-parity violating SUSY. Unlike in the MSSM, here, one can generate the light neutrino masses and mixing angles without adding any new particles to the MSSM particle content. Neutrino mass generation can be achieved through both spontaneous \cite{Dawson:1985vr,Romao:1991ex,Nogueira:1990wz,Romao:1991tp,Giudice:1992jg,Kitano:1999qb,Frank:2007un} and explicit \cite{Dawson:1985vr,Hall:1983id,Lee:1984kr,Lee:1984tn,Dimopoulos:1988jw,Diaz:1997xc,Joshipura:1994ib,deCampos:1995ten,Banks:1995by,Hirsch:1998kc,Chang:1999nu,Datta:1999yd} breaking of R-parity. 

 %%%%%%%%%%%%%%%%%%%%%%%%%%%%%%%%%%%%%%%%%%%%%%%
\begin{figure}[h]
%\vspace{-0.45cm}
\centering
%      \begin{subfigure}[t]{0.45\textwidth}
    \includegraphics[width=0.4\textwidth]{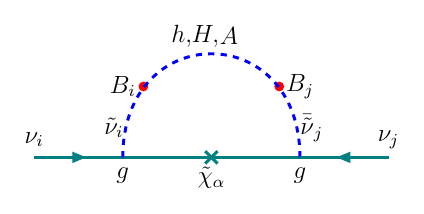}
    \hspace{0.7cm}
%    \caption{}
%    \end{subfigure}
%    \begin{subfigure}[t]{0.45\textwidth}
    \includegraphics[width=0.4\textwidth]{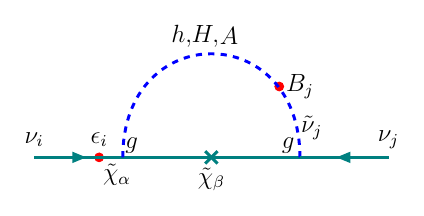}
%    \caption{}
%    \end{subfigure}
    \caption{One loop diagrams contributing to the neutrino mass: (Left) $BB$ loop and (Right)  $\epsilon B$ loop. For the $\epsilon B$ loop another contributing diagram is obtained by replacing $i \leftrightarrow j$.}
    \label{fig:loop}
\end{figure}
%%%%%%%%%%%%%%%%%%%%%%%%%%%%%%%%%%%%%%%%%%%%%%%%%

The R-parity can be spontaneously broken through left sneutrino VEV \cite{Aulakh:1982yn,Nieves:1983kn}. Through this mechanism, only one light neutrino mass can be generated at the tree level. The other light neutrino masses are generated at the loop level. A doublet Majoron is generated \cite{Nieves:1983kn} as a consequence of the sneutrinos acquiring VEVs. Such Majoron states are ruled out from electroweak precision data, e.g., Z-boson decay width \cite{ALEPH:1989kcj,OPAL:1989hoi,DELPHI:1989cmk} and astrophysical data \cite{Fukugita:1982ep,Kachelriess:2000qc}. This shortcoming can be overcome by adding a gauge singlet field with a lepton number \cite{Masiero:1990uj,Romao:1992vu}. The lepton number is broken by giving VEV to this gauge singlet state \cite{Masiero:1990uj,Romao:1992vu}. Phenomenological implications of these models have been studied in detail \cite{Hirsch:2004rw,Hirsch:2006di,Bhattacharyya:2010kr}.  

We now move to the RPV scenario that is more attractive phenomenologically and also widely studied, i.e., the explicit R-parity breaking scenario. As already discussed, there can be three types of lepton number violating terms and one baryon number violating term that can break R-parity explicitly. Among these terms, the bilinear RPV terms can generate light neutrino masses at the tree level. Trilinear couplings $\lambda$ and $\lambda^{\prime}$ can contribute to the neutrino masses at one-loop level. Note that the bilinear RPV term also induces correction to the tree-level masses of the neutrinos at the one-loop level. 

Let us address the tree-level contribution first before moving to loop contributions. Bilinear R-parity violating term induces a mixing between the light neutrino and neutralino states \cite{Barbier:2004ez,Dreiner:1997uz,Banks:1995by}. More precisely, the bilinear RPV term in the superpotential ($\epsilon_i \hat{L}_i \hat{H}_u$ where $i=1,2,3$) gives rise to mixing between up-type Higgsino ($\tilde H_u^0$) and three light neutrinos ($\nu_{iL}$). At the tree level, we now have a $7\times 7$ neutralino-neutrino mass matrix in the basis 
$\psi^0 =\left( \begin{matrix}
\tilde{B} & \tilde{W}_3 &\tilde{H}_d^0 &\tilde{H}_u^0 &\nu_e &\nu_{\mu} &\nu_{\tau} 
\end{matrix}\right )$ \cite{Banks:1995by, Nardi:1996iy, Grossman:1998py, Rakshit:2004rj}.  
%\begin{equation}
%\label{eq:matrix1}
%\begin{pmatrix}
%M_1 & 0 & -\frac{1}{2}g^{\prime} v_d & -\frac{1}{2} g^{\prime}v_u & -\frac{1}{2}g^{\prime} v_1 & -\frac{1}{2}g^{\prime}v_2 & -\frac{1}{2}g^{\prime}v_3 \\
%
%0 & M_2 & \frac{1}{2}gv_d & -\frac{1}{2}gv_u & \frac{1}{2}gv_1 & \frac{1}{2}gv_2 & \frac{1}{2}gv_3 \\
%
%-\frac{1}{2}g^{\prime}v_d & \frac{1}{2}gv_d & 0 & -\mu & 0 & 0 & 0 \\
%
%\frac{1}{2}g^{\prime}v_u & -\frac{1}{2}gv_u & -\mu & 0 & \epsilon_1 & \epsilon_2 & \epsilon_3 \\
%
%-\frac{1}{2}g^{\prime}v_1 & \frac{1}{2}gv_1 & 0 & \epsilon_1 & 0 & 0 & 0 \\
%
%-\frac{1}{2}g^{\prime}v_2 & \frac{1}{2}gv_2 & 0 & \epsilon_2 & 0 & 0 & 0 \\
%
%-\frac{1}{2}g^{\prime}v_3 & \frac{1}{2}gv_3 & 0 & \epsilon_3 & 0 & 0 & 0 
%\end{pmatrix}
%\end{equation}
%Here $\tilde{B}$ and $\tilde{W}_3$ represent bino and wino respectively while $M_1$ are $M_2$ are bino and wino mass parameters. The up-type and down-type Higgs boson VEVs are denoted by $v_u$ and $v_d$ respectively while $\langle\tilde{\nu}_i\rangle \equiv v_i (i = 1,2,3)$ represent the sneutrino VEVs. 
Diagonalization of this mass matrix results in one of the light neutrinos gaining non-zero mass apart from the four massive neutralinos.

The other neutrino masses are generated at the one-loop level in the bilinear RPV framework. The neutrino mass generated at the tree level also gets some correction at one loop. Two kinds of loop diagrams affect light neutrino masses in this model. They are referred to as $BB$ loop and $\epsilon B$ loop \cite{Davidson:2000uc,Davidson:2000ne}. 

Fig.~\ref{fig:loop} shows the Feynman diagram corresponding to these two loops. Fig.\ref{fig:loop}(a) refers to the $BB$ loop where the red dot ($B_i$, $B_j$) indicates mixing between the neutral Higgs states ($h,H,A$) and the sneutrinos. Fig~\ref{fig:loop}(b) refers to $\epsilon B$ loop where the red dots marked with ${\epsilon_{i}}$ and $B_j$ indicate mixing between the neutrino and the neutralino states and Higgs and the sneutrino states respectively. We get another possible loop diagram for the $\epsilon B$ loop if we make the replacement $i \leftrightarrow j$. 

After combining both the tree level and one-loop contributions to the neutrino mass within the bilinear RPV framework, one can write the light $3\times 3$ neutrino mass matrix as \cite{Rakshit:2004rj,Barbier:2004ez,Choudhury:2023lbp}
\begin{equation}
\begin{split}
\label{eq:total_mass}
[m_{\nu}]_{ij} & = [m_{\nu}]_{ij}^{\epsilon\epsilon} + [m_{\nu}]_{ij}^{BB} + [m_{\nu}]_{ij}^{\epsilon B} \\
& = X_T \epsilon_i \epsilon_j \sin^2\zeta + C_{ij} B_iB_j + (C_{ij}^\prime \epsilon_i B_j + i \leftrightarrow j) 
\end{split}
\end{equation}
Here $[m_{\nu}]_{ij}^{\epsilon\epsilon}$, $[m_{\nu}]_{ij}^{BB}$, $[m_{\nu}]_{ij}^{\epsilon B}$ represent the tree level, $BB$ loop and $\epsilon B$ loop contributions respectively. The alignment between $\epsilon_i$ and $v_i$\cite{Chun:2002vp,Grossman:2000ex,Nardi:1996iy,Borzumati:1996hd,Davidson:2000uc,Grossman:2003gq,Davidson:2000ne} is represented by the parameter $\zeta$. $X_T$ can be written as \cite{Rakshit:2004rj,Grossman:2003gq,Choudhury:2023lbp}
\beq
\label{eq:XT}
X_T = {m_Z^2 m_{\tilde \gamma}\cos^2\beta \over 
\mu(m_Z^2 m_{\tilde \gamma}\sin 2\beta-M_1 M_2 \mu)}
\eeq 
where $m_{\tilde \gamma}\equiv \cos^2\theta_w M_1 + \sin^2\theta_w M_2$. The $B_i$ parameters induce mixing between the sneutrino and the Higgs states. This results in finite mass splitting between CP-even and CP-odd sneutrino mass eigenstates. Non-zero neutrino masses generated from the $BB$ loop depend on this mass splitting \cite{Grossman:2003gq,Rakshit:2004rj,Davidson:2000uc}. This happens to be the more dominant contribution compared to that obtained from the $\epsilon B$ loop.  
The three individual contributions to the light $3\times 3$ neutrino mass matrix can be approximated through simple analytical expressions assuming that all the SUSY particle masses are at scale $\tilde m$ \cite{Grossman:2003gq,Choudhury:2023lbp}: 
\begin{eqnarray}\label{eq:tree_BB}
\begin{split}
[m_{\nu}]_{ij}^{\epsilon\epsilon} & \sim{\cos^2\beta\over \tilde m} \epsilon_i \epsilon_j \sin^2\zeta \\
[m_{\nu}]_{ij}^{BB} & \sim \frac{g^2}{64\pi^2\cos^2\beta} \frac{B_iB_j}{\tilde{m}^3} ~\epsilon_H \\ 
 [m_{\nu}]_{ij}^{\epsilon B} & \sim \frac{g^2}{64\pi^2\cos\beta} \frac{\epsilon_iB_j + \epsilon_jB_i}{\tilde{m}^2} ~\epsilon_H^\prime 
\end{split} 
\end{eqnarray}
where $\epsilon_H$ or $\epsilon_H^\prime$ appears in the expression because of the cancellation of contributions arising from different possible Higgs ($h,H,A$) diagrams in BB and $\epsilon$B loop. This gives rise to further suppression of the loop contributions to neutrino masses \cite{Rakshit:2004rj}. Apart from these two diagrams, one can have more non-trivial loop processes contributing to the neutrino masses \cite{Grossman:1997is,Davidson:2000uc,Davidson:2000ne,Grossman:2003gq}. 

\begin{figure}[h]
%\vspace{-0.45cm}
\centering
%      \begin{subfigure}[t]{0.45\textwidth}
    \includegraphics[width=0.4\textwidth]{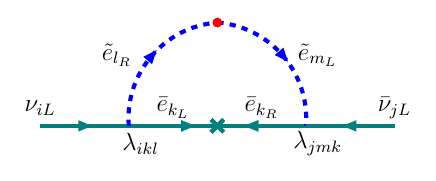}
    \hspace{0.7cm}
%    \caption{}
%    \end{subfigure}
%    \begin{subfigure}[t]{0.45\textwidth}
    \includegraphics[width=0.4\textwidth]{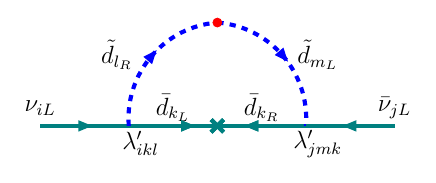}
%    \caption{}
%    \end{subfigure}
    \caption{One loop diagrams contributing to the neutrino mass: (Left) with $\lambda$ couplings and (Right) with $\lambda^{\prime}$ couplings. }
    \label{fig:trilin_neut}
\end{figure}
%%%%%%%%%%%%%%%%%%%%%%%%%%%%%%%%%%%%%%%%%%%%%%%%%

Now let us move to various loop contributions to the light neutrino masses arising from the lepton number violating trilinear terms ($\lambda$ and $\lambda^{\prime}$). These contributions are the leading ones only when the bilinear couplings are either absent or very highly suppressed.   

The dominant contributions arise from the diagrams shown in Fig.~\ref{fig:trilin_neut}. For this class of diagrams, we always require two trilinear couplings since the resultant neutrino mass term is Majorana type, which violates the lepton number by two units. As evident from the figure, the $\lambda$ and $\lambda^{\prime}$ couplings contribute to the neutrino mass matrix through lepton-slepton and quark-squark loops \cite{Hall:1983id, Babu:1989px, Barbier:2004ez}.  
\begin{equation}
\begin{split}
[m_{\nu}]_{ij}^{\lambda\lambda} & = \frac{1}{16\pi^2}\sum_{k,l,m} \lambda_{ikl}\lambda_{jmk}{m_e}_k\frac{(\tilde{m}_{LR}^{e2})_{ml}}{m^2_{\tilde{e}_{Rl}}-m^2_{\tilde{e}_{Lm}}}{\rm ln}\left( \frac{m^2_{\tilde{e}_{Rl}}}{m^2_{\tilde{e}_{Lm}}} \right) + (i \leftrightarrow j) \\
m_{\nu}]_{ij}^{\lambda^{\prime}\lambda^{\prime}} & = \frac{3}{16\pi^2}\sum_{k,l,m} \lambda^{\prime}_{ikl}\lambda^{\prime}_{jmk}{m_d}_k \frac{(\tilde{m}_{LR}^{d2})_{ml}}{m^2_{\tilde{d}_{Rl}}-m^2_{\tilde{d}_{Lm}}}{\rm ln}\left( \frac{m^2_{\tilde{d}_{Rl}}}{m^2_{\tilde{d}_{Lm}}} \right) + (i \leftrightarrow j) 
\end{split}
\end{equation}
In order to write this expression, it is assumed that the left-right slepton mass squared matrix $\tilde{m}_{LR}^{e2}$ and the left-right squark mixing matrix $\tilde{m}_{LR}^{d2}$ are written in a basis where the charged lepton masses and the down quark masses are diagonal. 
A class of loop diagrams can generate neutrino mass in the presence of both bilinear and trilinear RPV couplings \cite{Davidson:2000uc,Davidson:2000ne,Grossman:2003gq,Rakshit:2004rj}. These contributions are proportional to either $\epsilon\lambda$ or $\epsilon\lambda^{\prime}$. Apart from the loop factor, they are further suppressed by a fermion mass \cite{Grossman:2003gq,Rakshit:2004rj}.
%%%%%%%%%%%%%%%%%%%%%%%%%%%%%%%%%%%%%%%%%%%%%%%
\begin{figure}[!htpb]
%    \begin{subfigure}[t]{0.49\textwidth}
    \includegraphics[width=0.45\textwidth]{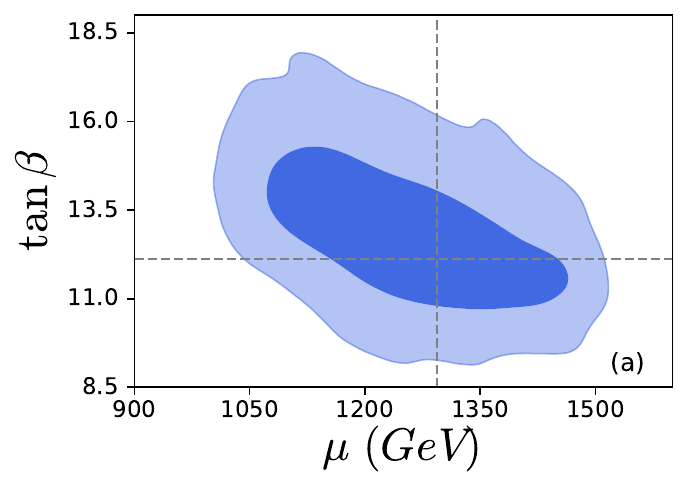}
   % \caption{}
%    \label{fig:nh_eps12}
%    \end{subfigure}
%    \begin{subfigure}[t]{0.49\textwidth}
    \includegraphics[width=0.45\textwidth]{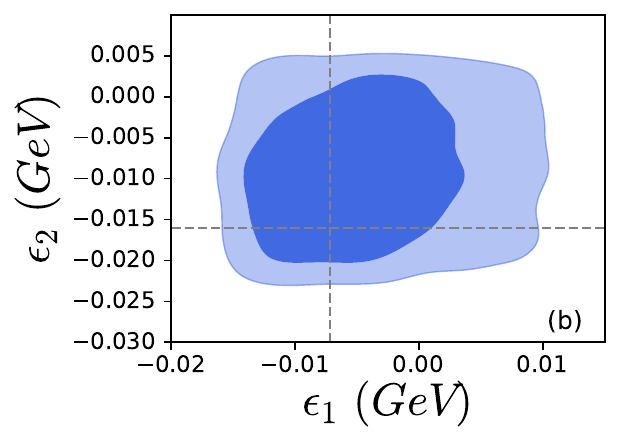}
%    \end{subfigure}
    \\[\smallskipamount]
%    \begin{subfigure}[t]{0.5\textwidth}
    \includegraphics[width=0.45\textwidth]{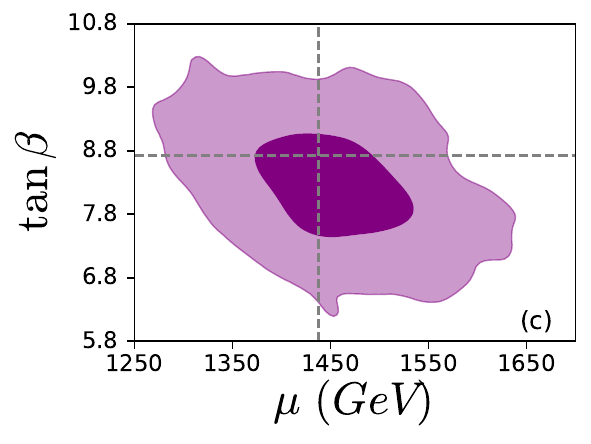}
%    \end{subfigure}
%    \hfill
%    \begin{subfigure}[t]{0.48\textwidth}
    \includegraphics[width=0.45\textwidth]{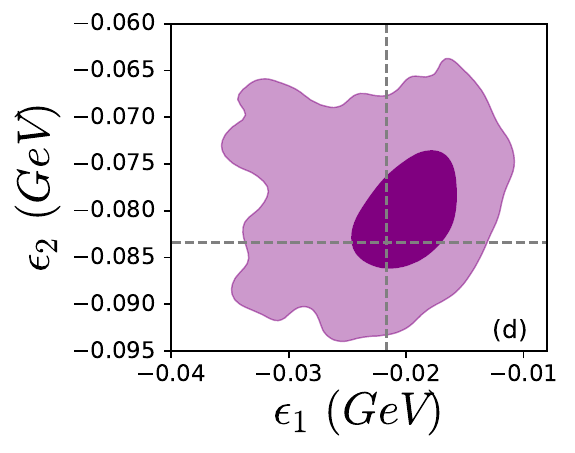}
%    \end{subfigure}
    \caption{Marginalized distributions with $68\%$ (dark blue/magenta) and $95\%$ (light blue/magenta) confidence level contours in the (a) $\mu$-${\rm tan}\beta$ plane and (b) $\epsilon_1$-$\epsilon_2$ plane for the NH case and (c) $\mu$-${\rm tan}\beta$ plane and (d) $\epsilon_1$-$\epsilon_2$ plane for the IH case \cite{Choudhury:2023lbp}.}
    \label{fig:nh_bilin}
\end{figure}
%%%%%%%%%%%%%%%%%%%%%%%%%%%%%%%%%%%%%%%%%%%%%%%%%

It is evident that given the precision of neutrino oscillation data at present, one can very effectively constrain the RPV couplings relevant to the neutrino mass calculation as well as the SUSY particle masses appearing at the loop. One such study was performed recently in the context of bilinear R-parity violation \cite{Choudhury:2023lbp}. In this study, the interplay among the RPV parameters, namely, bilinear couplings ($\epsilon_i$), their corresponding soft terms ($B_i$), and the sneutrino VEVs ($v_i$) was studied from the perspective of neutrino mass, Higgs mass and couplings measurements and flavor observables. Both normal hierarchy (NH) and inverted hierarchy (IH) scenarios were considered while fitting the oscillation data. This study was performed with 9 above-mentioned RPV parameters and two MSSM parameters, namely, $\mu$ and $\tan\beta$. Altogether, 15 observables were included: two neutrino mass squared differences, three mixing angles, the SM Higgs boson mass and its seven coupling strength modifiers, and two flavor constraints arising from b-decay. The parameter space was explored by adopting the MCMC algorithm through the \texttt{emcee} package \cite{Foreman-Mackey_2013}. The objective was not only to find the best-fit parameter values but also the $68
\%$ and $95\%$ confidence level regions of the 9 RPV parameters alongside $\mu$ and ${\rm tan}\beta$.  

The results were presented in the form of two-dimensional marginalized distributions, which indicated the $1\sigma$ and $2\sigma$ allowed regions of the RPV parameters \cite{Choudhury:2023lbp}. Fig.~\ref{fig:nh_bilin} showcases some results derived assuming normal and inverted mass hierarchies of the light neutrinos. $\chi^2_{min}/DoF= 0.865$ and $0.845$ were obtained in the NH and IH scenarios respectively. The results clearly show that the RPV parameters $\epsilon_i$ and $v_i$ have to be quite small. The $\epsilon_i$ and $v_i$ parameters are constrained to be $|\epsilon_i|\sim 10^{-2} - 10^{-3}$ and $v_i\sim 10^{-4}$ GeV in the NH scenario. For the IH scenario, the constraints obtained were as follows: $|\epsilon_i|\sim 10^{-2} - 10^{-3}$ and $v_i\sim 10^{-3} - 10^{-4}$. The $B_i$ parameters only contribute to the loop. Given the hierarchical structure, the IH case typically requires larger loop corrections to all three neutrino states leading to larger $B_i$ values. The neutrino mass hierarchy is reflected in the resultant ranges obtained for all $\epsilon_i$, $v_i$ and $B_i$ parameters\footnote{For more detailed results, please refer to \cite{Choudhury:2023lbp}.}. $\mu$ and $\tan\beta$ were also very highly constrained in the process. The typical allowed ranges for $\mu$ and $\tan\beta$ are $\sim$ 1-1.5 (1.3-1.6) TeV and $\sim$ 8-16 (6-10) for NH (IH) scenario, respectively.

These results are very significant since any BSM theory has to be able to explain the non-zero light neutrino mass and mixing phenomena, and therefore, one has to take into account the constraints derived in this article while performing any phenomenological studies with the framework of bilinear RPV. This work is the only one available in the literature that explicitly derives the boundaries on the bilinear RPV couplings from neutrino oscillation and 125 GeV Higgs data. A similar study with the trilinear couplings is lacking in the existing literature.  
%%%%%%%%%%%%%%%%%%%%%%%%%%%%%%%%%%%%%%%%%%%%%%%%%%%%%%%%%%%%%%%%%%%%%%%%%%%%%%%%%%%%%%%%%%%%%%%%%%%%%%%% 
%%%%%%%%%%%%%%%%%%%%%%%%%%%%%%%%%%%%%%%%%%%%%%%%%%%%%%%%%%%%%%%%%%%%%%%%%%%%%%%%%%%%%%%%%%%%%%%%%%%%%%%%
\section{Anomalous muon magnetic moment}
\label{sec:mug2}
%%%%%%%%%%%%%%%%%%%%%%%%%%%%%%%%%%%%%%
%%%%%%%%%%%%%%%%%%%%%%%%%%%%%%%%%%%%%%%%%%%%%%%
\begin{figure}[h]
%\vspace{-0.45cm}
\centering
      \includegraphics[width=0.4\textwidth]{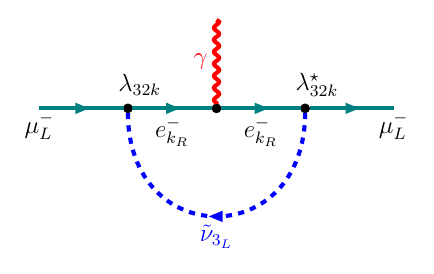} \hspace{+0.5cm}
      \includegraphics[width=0.4\textwidth]{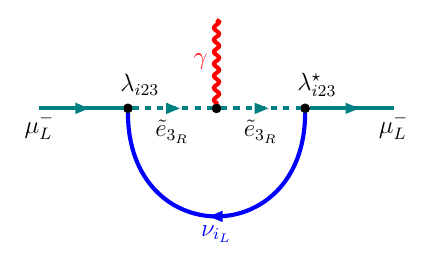}
      \includegraphics[width=0.4\textwidth]{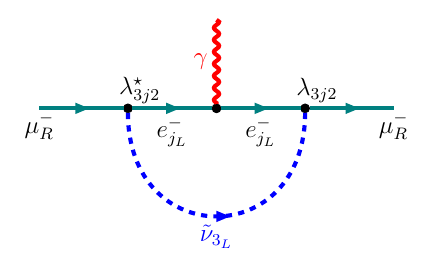}
 \hspace{+0.5cm}
      \includegraphics[width=0.4\textwidth]{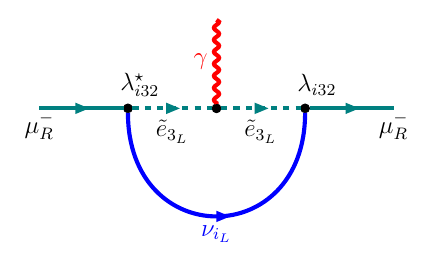}
     \caption{One loop diagrams contributing to the muon (g-2) with non-zero $\lambda$ couplings. }
    \label{fig:mug2_lam}
\end{figure}
%%%%%%%%%%%%%%%%%%%%%%%%%%%%%%%%%%%%%%%%%%%%%%%%% 
A recent measurement \cite{Muong-2:2023cdq} by the Muon (g-2) collaboration at Fermilab puts the existing anomaly in $\Delta a_{\mu}$ \cite{Muong-2:2006rrc,Muong-2:2021vma,Muong-2:2021ojo,Muong-2:2023cdq} more than $5\sigma$ standard deviation away from the SM prediction \cite{Aoyama:2020ynm,Cvetic:2020unz}. Given the existing collider limits on the electroweak sector particles of RPC MSSM, there only remains a very small parameter space where the anomaly can be explained \cite{Chakraborti:2015mra,Chakraborti:2021bmv,Chakraborti:2021dli,Endo:2021zal}. RPV MSSM provides an additional contribution to    $\Delta a_{\mu}$ through different non-zero couplings, thereby making it a more attractive scenario to fit this particular observation. The effect of the bilinear RPV couplings on the RPC MSSM contribution does not give rise to any new class of contributing diagrams. The major contribution still arises from neutralino-slepton and chargino-sneutrino loops. However, owing to the mixing between the chargino-charged lepton, neutralino - neutrino states, sneutrino - neutral Higgs, and slepton-charged Higgs; there are now additional loop diagrams that contribute to muon (g-2). A generic expression arising from these two loops can be written as \cite{Martin:2001st,Moroi:1995yh,Hundi:2011si}
\begin{equation}
\begin{split}
\Delta a_{\mu}^{N^0\tilde\mu} & = \frac{m_{\mu}}{16\pi^2} \sum_{A,j}\{-\frac{m_{\mu}}{6m^2_{\tilde\mu A}(1-x_{Aj})^4} (N^L_{Aj}N^L_{Aj}+N^R_{Aj}N^R_{Aj})(1-6x_{Aj}+3x^2_{Aj} \\  & +2x^3_{Aj}-6x^2_{Aj}{\rm ln}x_{Aj}) -\frac{(m_{N^0})_j}{m^2_{\tilde\mu A}(1-x_{Aj})^3}N^L_{Aj}N^R_{Aj}(1-x^2_{Aj}+2x_{Aj}{\rm ln}x_{Aj}) \} \\
\Delta a_{\mu}^{C^{\pm}\tilde\nu_{\mu}} & = \frac{m_{\mu}}{16\pi^2} \sum_{j}\{\frac{m_{\mu}}{6m^2_{\tilde\nu_{\mu}}(1-x_j)^4} (C^L_jC^L_j + C^R_jC^R_j)(2+3x_j-6x^2_j+x^3_j \\ & +6x_j{\rm ln}x_j)  -\frac{(m_{C^{\pm}})_j}{m^2_{\tilde\nu_{\mu}}(1-x_j)^3}C^L_jC^R_j(3-4x_j+x_j^2+2{\rm ln}x_j) \} 
\end{split}
\end{equation} 
Here 
\begin{equation}
\begin{split}
x_{Aj} & = \frac{(m^2_{N^0})_j}{m^2_{\tilde\mu A}}, ~~x_j = \frac{(m^2_{C^{\pm}})_j}{m^2_{\tilde\mu}}, \\
N^L_{Aj} & = -y_{\mu}(U_0)_{4j}(U_{\tilde\mu})_{LA} - \sqrt{2}g_1(U_0)_{1j}(U_{\tilde\mu})_{RA}, \\
N^R_{Aj} & = -y_{\mu}(U_0)_{4j}(U_{\tilde\mu})_{RA} + \frac{1}{\sqrt{2}} (g_2(U_0)_{2j} + g_1(U_0)_{1j})(U_{\tilde\mu})_{LA}, \\
C^L_j & = y_{\mu}(U_-)_{2j}, ~~C^R_j = -g_2(U_+)_{1j}.
\end{split}
\end{equation}
%%%%%%%%%%%%%%%%%%%%%%%%%%%%%%%%%%%%%%%%%%%%%%%
\begin{figure}[h]
%\vspace{-0.45cm}
\centering
      \includegraphics[width=0.4\textwidth]{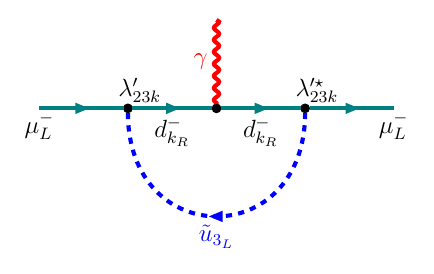}
      \hspace{+0.5cm}
      \includegraphics[width=0.4\textwidth]{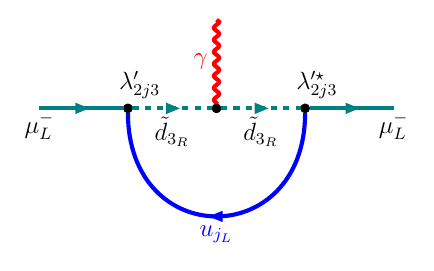}
      \includegraphics[width=0.4\textwidth]{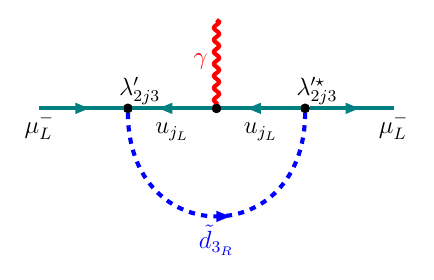}
      \hspace{+0.5cm}
      \includegraphics[width=0.4\textwidth]{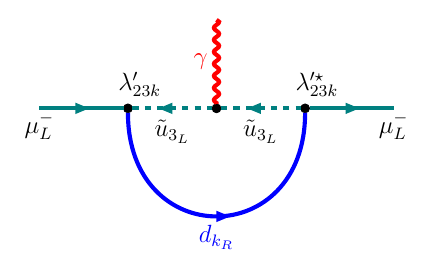}
     \caption{One loop diagrams contributing to the muon (g-2) with non-zero $\lambda^{\prime}$ couplings. }
    \label{fig:mug2_lamp}
\end{figure}
%%%%%%%%%%%%%%%%%%%%%%%%%%%%%%%%%%%%%%%%%%%%%%%%%  
Muon mass and Yukawa coupling are represented by $m_{mu}$ and $y_{\mu}$. Light neutrino mass matrix is diagonalised by $U_0$. $U_{\pm}$ diagonalised the chargino mass matrix. The gaugino and higgsino mass parameters, along with the slepton masses, are crucial in calculating the $\Delta a_{\mu}$ contribution. ${\rm tan}\beta$ also plays an important role. Apart from these, the RPV parameters $\epsilon_i$, $B_i$, and $v_i$, which induce the mixing between the sparticle-particle pairs mentioned above, are vital. However, if one attempts to fit the neutrino oscillation data as well as muon (g-2) excess within the bilinear RPV framework, the RPV contribution to  $\Delta a_{\mu}$ is rendered small due to the smallness of the RPV parameters which is a result of solar and atmospheric neutrino mass scales \cite{Choudhury:2023lbp}. 

Among the trilinear couplings, the lepton number violating ones, namely, $\lambda$ and $\lambda^{\prime}$ can, in principle, contribute to muon (g-2) \cite{Kim:2001se,Chakraborty:2015bsk}. However, the $\lambda$ contribution will dominate since in the presence of $\lambda^{\prime}$ couplings, the loops that contribute to muon (g-2) have squarks appearing in the propagator as opposed to sleptons and sneutrinos in the other case. Given the present collider constraints, the squarks, in general, have to be much heavier than the slepton/sneutrino masses. The most dominant contribution with non-zero $\lambda$ couplings arises from the diagrams shown in Fig.~\ref{fig:mug2_lam}. 

The collider constraints on the first two generation sleptons and sneutrinos are stronger compared to that on the third generation. There are further constraints on the first two generation right chiral sleptons arising from neutrino physics as well as other low energy observables. The generic simplified expression for $\lambda$ contribution to $\Delta a_{\mu}$ therefore can be written as \cite{Kim:2001se,Chakraborty:2015bsk}
\bea
[\Delta a_{\mu}]^{\lambda\lambda} = \frac{m^2_{\mu}}{96\pi^2}[ |\lambda_{23k}|^2\frac{2}{m^2_{\tilde{\nu}_{\tau}}} + |\lambda_{3k2}|^2\{ \frac{2}{m^2_{\tilde{\nu}_{\tau}}} - \frac{1}{m^2_{\tilde{\tau}_L}}\} - |\lambda_{k23}|^2 \frac{1}{m^2_{\tilde{\tau}_R}}  ]
\eea
Here, $m_{\tilde{\tau}_L}$, $m_{\tilde{\tau}_R}$ and $m_{\tilde{\nu}_{\tau}}$ represent the left stau, right stau, and tau sneutrino masses respectively. Similarly, one can obtain a contribution to muon (g-2) when the $\lambda^{\prime}$ couplings are non-zero from the kind of diagrams shown in Fig.~\ref{fig:mug2_lamp}. 

%%%%%%%%%%%%%%%%%%%%%%%%%%%%%%
\section{Flavor Observables}
\label{sec:flavor}
\begin{figure}[h]
%\vspace{-0.45cm}
\centering
      \includegraphics[width=0.4\textwidth]{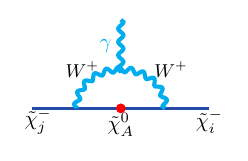} \hspace{0.7cm}
      \includegraphics[width=0.4\textwidth]{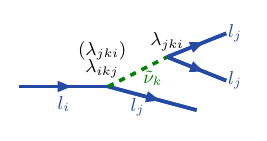}
     \caption{Diagrams contributing to $l_j\to l_i\gamma$ and $l_i\to l_jl_jl_j$ with non-zero bilinear and trilinear $\lambda$ couplings respectively.}
    \label{fig:lep_vio_rpv}
\end{figure}

%%%%%%%%%%%%%%%%%%%%%%%%%%%%%%
There are multiple sources for flavor violation both in the lepton and quark sectors within the RPV MSSM framework. Lepton flavor violating (LFV) decays and resulting constraints have been studied extensively in the context of bilinear and trilinear ($\lambda$ and $\lambda^{\prime}$) couplings have been studied extensively \cite{Cheung:2001sb, Carvalho:2002bq, Vicente:2013fya, Choi:2000bm, deGouvea:2000cf, Gemintern:2003gd, Chen:2008gd, Arhrib:2012ax, Arhrib:2012mg, Dreiner:2012mx, Bose:2010eb}. Fig.~\ref{fig:lep_vio_rpv} shows some sample Feynman diagrams that can contribute to several LFV processes through bilinear/trilinear couplings. 
%%%%%%%%%%%%%%%%%%%%%%%%%%%%%%%%%%%%%%%%%%%%%%%
%%%%%%%%%%%%%%%%%%%%%%%%%%%%%%%%%%%%%%%%%%%%%%%%%
On the left of Fig.~\ref{fig:lep_vio_rpv} bilinear contribution to decays like $l_j\to l_i\gamma$ is depicted. Owing to the bilinear RPV term in the superpotential, mixing is generated among the charged leptons-charginos and neutrino-neutralinos. As a result, there are multiple subprocesses that can contribute to the above-mentioned decays. Non-observation of such decays puts constraints on the relevant couplings. On the right of Fig.~\ref{fig:lep_vio_rpv} one example diagram is shown which directly contributes to LFV decays like $l_i\to l_jl_jl_j$. Trilinear $\lambda$ couplings can be constrained severely from the non-observation of such decays. Refer to \cite{deGouvea:2000cf} for such constrints. Updated studies in this context are lacking in the literature.   

In the quark sector existing studies have explored meson mixing observables such as $\Delta M_s$, $\Delta M_d$ and $\Delta M_K$ corresponding to $B_s^0$, $B_d^0$ and $K^0$ mesons respectively \cite{Gabbiani:1996hi, Altmannshofer:2007cs, Domingo:2018qfg}. At the tree level, the simplest process that can contribute to this kind of mass difference observables involves quark scattering mediated by sneutrinos or squarks. After taking into account the due one loop corrections, one can derive bounds on the $\lambda^{\prime}$ and $\lambda^{\prime\prime}$ couplings subjected to the sneutrino and squark masses respectively, for example, \cite{Domingo:2018qfg}
\begin{equation}
\begin{split}
\lambda^{\prime}_{i13}\lambda^{\prime}_{i31} &\lesssim 1.6\times 10^{-6}\left( \frac{m_{\tilde\nu_i}}{1{~\rm TeV}} \right)^2 \\
|\lambda^{\prime\prime}_{112}\lambda^{\prime\prime}_{123}| &\lesssim 2.8\times 10^{-2} \left( \frac{m_{\tilde s_R, \tilde u_R}}{1{~\rm TeV}} \right)^2
\end{split}
\end{equation}   

%%%%%%%%%%%%%%%%%%%%%%%%%%%%%%%%%%%%%%%%%%%%%%%
\begin{figure}[h]
%\vspace{-0.45cm}
\centering
      \includegraphics[width=0.4\textwidth]{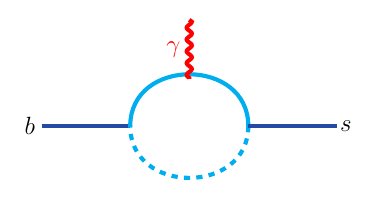} \hspace{0.7cm}
      \includegraphics[width=0.4\textwidth]{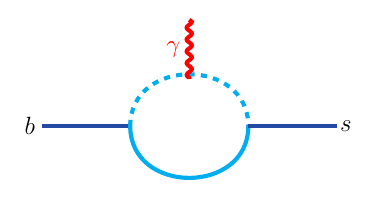}
     \caption{Diagrams contributing to $b\to s\gamma$ with non-zero trilinear RPV couplings.}
    \label{fig:flav_b2sg}
\end{figure}
%%%%%%%%%%%%%%%%%%%%%%%%%%%%%%%%%%%%%%%%%%%%%%%%%
Apart from that observables from leptonic and hadronic B-decays have also been explored in the context of RPV SUSY \cite{deCarlos:1996yh, Dreiner:2001kc, Dreiner:2013jta}. Contribution to the decay $b\to s\gamma$ can arise from either $\lambda^{\prime}$ or $\lambda^{\prime\prime}$ couplings as shown in Fig.~\ref{fig:flav_b2sg}.
A combination $\lambda_{ij2}^{\prime}\lambda_{ij3}^{\prime} \neq 0$ or $\lambda_{i12}^{\prime\prime}\lambda_{i13}^{\prime\prime} \neq 0$ can contribute to the decay width.  

\begin{figure}[htb]
%\vspace{-0.45cm}
\centering
      \includegraphics[width=0.3\textwidth]{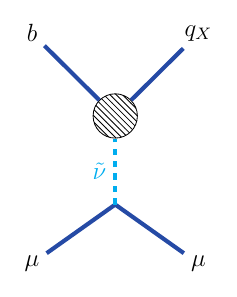}
      \hspace{1.0cm}
      \includegraphics[width=0.5\textwidth]{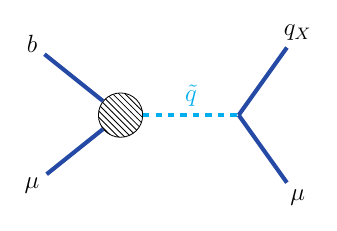}
     \caption{Diagrams contributing to $B_{s,d}\to \mu^+\mu^-$ with non-zero $\lambda^{\prime}$ couplings.}
    \label{fig:flav_bs2mm}
\end{figure}

\begin{figure}[htb]
%\vspace{-0.45cm}
\centering
      \includegraphics[width=0.3364\textwidth]{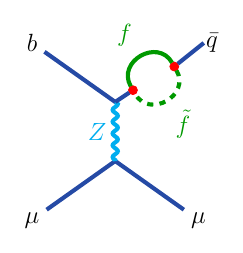}
      \hspace{+1.0cm}
      \includegraphics[width=0.3364\textwidth]{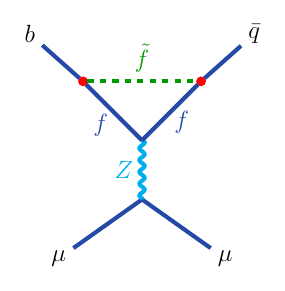}
     \caption{Diagrams contributing at one loop to $B_{s,d}\to \mu^+\mu^-$ with non-zero $\lambda^{\prime}$ and $\lambda^{\prime\prime}$ couplings \cite{Dreiner:2013jta}.}
    \label{fig:flav_bs2mm_loop}
\end{figure}

There are even more possibilities if one considers $B_{s,d}\to \mu^+\mu^-$. There can be contribution at the tree level itself through non-zero $\lambda$ and $\lambda^{\prime}$ couplings as shown in Fig.~\ref{fig:flav_bs2mm}. 
%%%%%%%%%%%%%%%%%%%%%%%%%%%%%%%%%%%%%%%%%%%%%%%
%%%%%%%%%%%%%%%%%%%%%%%%%%%%%%%%%%%%%%%%%%%%%%%%%
Following this one can derive bounds on the RPV couplings subjected to the sparticle masses \cite{Dreiner:2013jta}. 
\begin{equation}
\begin{split}
|\lambda_{i22}\lambda^{\prime}_{i12}| &< 2.2\times 10^{-7}[m^2_{\tilde{\nu}_L}]  \\
|\lambda^{\prime}_{2i2}\lambda^{\prime}_{2i2}| &< 8.2\times 10^{-5}[m^2_{\tilde{u}_i}]
\end{split}
\end{equation}
At one loop level both $\lambda^{\prime}$ and $\lambda^{\prime\prime}$ couplings can contribute to the decay as shown in Fig.~\ref{fig:flav_bs2mm_loop}. 
%%%%%%%%%%%%%%%%%%%%%%%%%%%%%%%%%%%%%%%%%%%%%%%
%%%%%%%%%%%%%%%%%%%%%%%%%%%%%%%%%%%%%%%%%%%%%%%%%
For more details, refer to \cite{Dreiner:2013jta}.
%%%%%%%%%%%%%%%%%%%%%%%%%%%%
\section{Collider analysis}
\label{sec:collider}
%%%%%%%%%%%%%%%%%%%%%%%%%%%%
The large number of RPV couplings gives rise to a plethora of different final states, which have been studied by the ATLAS and CMS collaborations over the years. The exclusion limits on the SUSY particles vary widely in the RPV framework compared to those in the RPC framework. In this section, we summarise the existing results first for different non-zero trilinear couplings and then for non-zero bilinear couplings. While doing that, we mostly concentrate on RUN-II data, which provides the most updated and stringent exclusion limits thus far. Before concluding 
this section we will summarize a few recent works where the search prospects of the electroweakinos have been explored in the context of RPV SUSY at the HL-LHC and/or 
HE-LHC.

%%%%%%%%%%%%%%%%%%%%%%%%%%%%%%%%%%%%
\subsection{Gluino search status}
\label{sec:gluon}
%%%%%%%%%%%%%%%%%%%%%%%%%%%%%%%%%%%%

The most stringent bound has been obtained for gluino pair production 
where both direct and cascade decay of the gluinos into the SM particles have been studied. 
Different decay modes result in different final states at the LHC. Here, we categorize the gluino searches depending on the final states explored by the ATLAS and CMS collaborations.   

\begin{itemize}

\item \textbf{Same sign dileptons or three leptons and multi jets:} 
In this search, two leptons with the same sign charge or three leptons with multiple jets are considered in the final state. The final state is also categorized by the presence of the number of b-jets and threshold of the effective mass, $m_{eff}$\footnote{The effective mass is defined as $m_{eff}=\sum_i p_T^{l_i} + \sum_i p_T^{j_i} + \met$}. The ATLAS collaboration considered RIN-II data to put bounds on gluino mass in these signal regions \cite{ATLAS:2023afl, ATLAS:2019fag, ATLAS:2017tmw}. These signal regions can arise from either non-zero $\lambda^{\prime}$ or $\lambda^{\prime\prime}$ couplings. The mass of the gluino is excluded up to 2.2 TeV for $m_{\lspone} <$ 1000 GeV from this search \cite{ATLAS:2023afl}. A similar analysis has been performed by the CMS collaboration, and they exclude $m_{\tilde{g}}$ below 2.1 TeV as shown on the middle panel of Fig.~\ref{fig:gluino_reach} \cite{CMS:2020cpy}.  

\item \textbf{Four or more leptons:}
This kind of multilepton final state can arise when both the LSP ($\lspone$) decay via $\lambda_{ijk}$ type coupling. There are several searches done by ATLAS and CMS corresponding to this final state~\cite{ATLAS:2021yyr, ATLAS:2018rns}. In this kind of search, the events with $N_l(l=e,\mu) \geq $ 4 are considered. Here, two different scenarios are considered corresponding to different couplings like $\lambda_{12k}$ with $k \equiv 1, 2$ and $\lambda_{i33}$ with $i \equiv 1, 2$.  
From RUN-II data, the gluino mass is excluded up to 2.5 TeV and 2.0 TeV when non-zero $\lambda_{12k}$ and $\lambda_{i33}$ coupling are considered respectively as shown on the left of Fig.~\ref{fig:gluino_reach} \cite{ATLAS:2021yyr}. 

%%%%%%%%%%%%%%%%%%%%%%%%%%%%%%%%
\begin{figure}[!htb]
\begin{center}
\includegraphics[width=4.5cm,height=4.25cm]{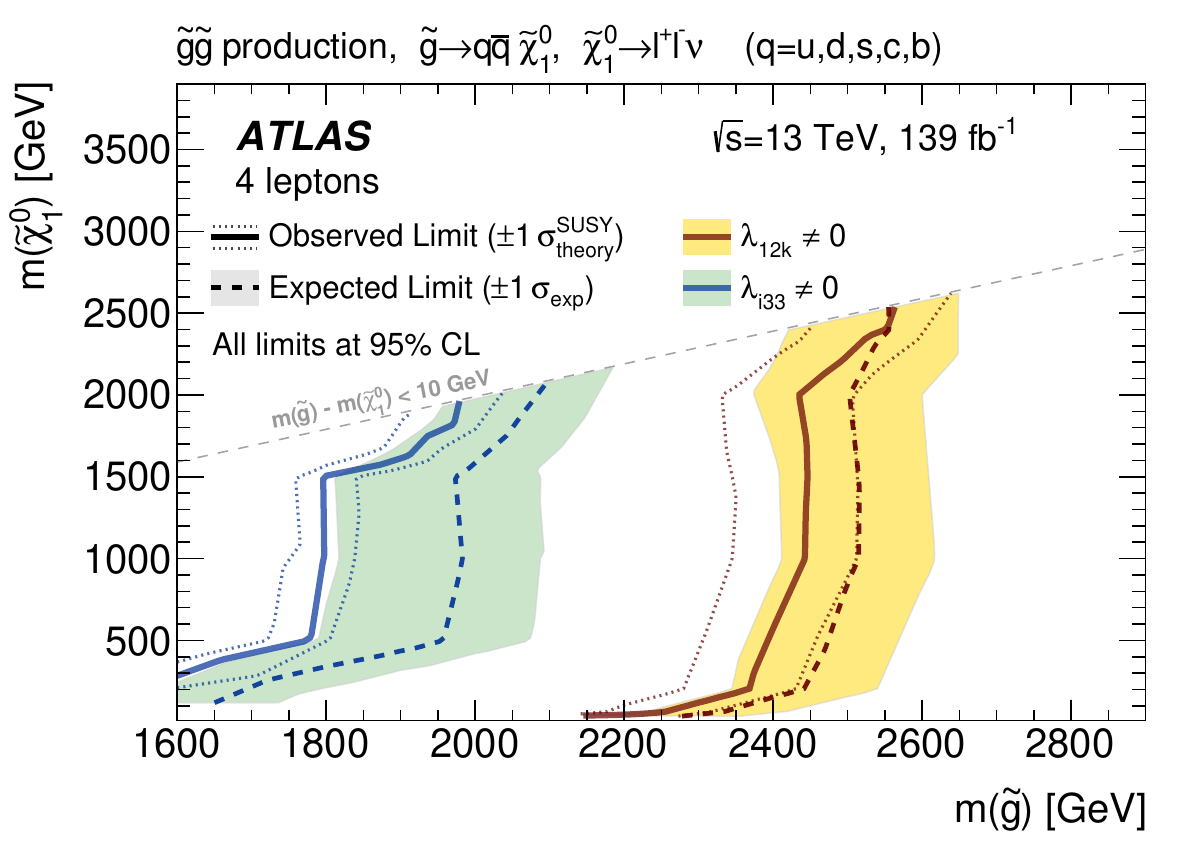}
\includegraphics[width=4.5cm,height=4.25cm]{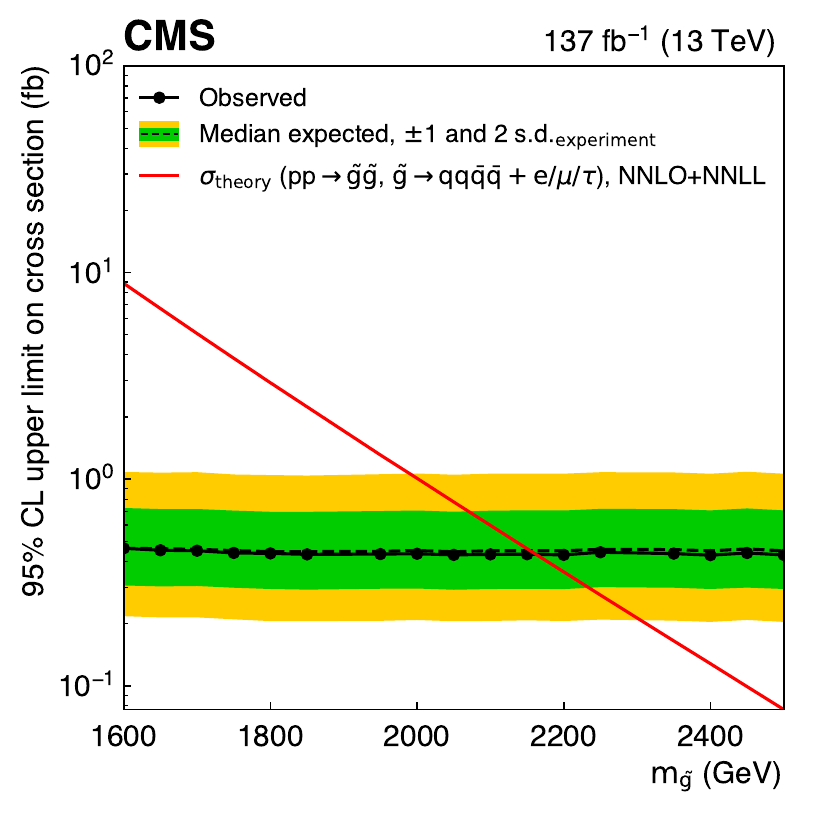}
\includegraphics[width=4.5cm,height=4.25cm]{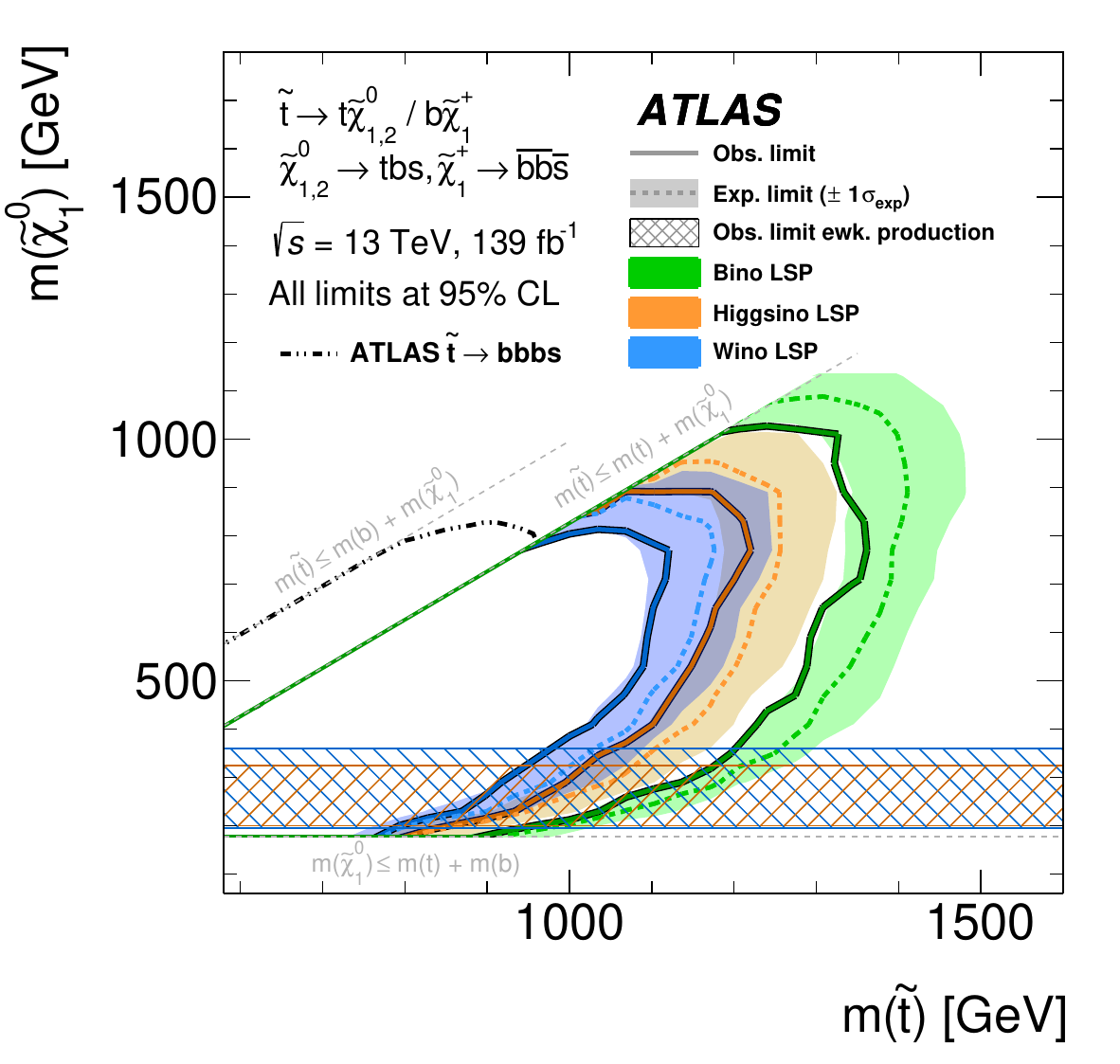}
\caption {The exclusion limits on gluino mass observed by (Left) the ATLAS collaboration in multi-lepton channel \cite{ATLAS:2021yyr} and (Middle) the CMS collaboration in multi-lepton multi-jet channel \cite{CMS:2020cpy} are shown. (Right) The exclusion limit on the lightest stop mass is shown as obtained by the ATLAS collaboration with final states containing at least one lepton and varied b-jet multiplicity \cite{ATLAS:2021fbt}.}
\label{fig:gluino_reach}
\end{center}
\end{figure}
%%%%%%%%%%%%%%%%%%%%%%%%%%%%%%%%%
\item \textbf{Multi jets:} 
Here, either the gluinos directly decay into quarks or decay into LSP, which eventually decays into the quarks via non-zero $\lambda^{\prime\prime}$ coupling. From cascade decay of gluino, $1000 < m_{\tilde{g}} < 1875$ GeV is excluded depending upon the choice of $m_{\lspone}$ \cite{ATLAS:2018umm}. For direct decay,  $m_{\tilde{g}} <$ 1.8 TeV is excluded, and for cascade decay, it is excluded up to 2.34 TeV for neutralino mass of 1.25 TeV \cite{atlas_web3}. A similar search performed by the CMS collaboration excludes gluino masses in the range 0.10 - 1.41 TeV \cite{CMS:2018pdq}. Similarly, a jet resonance search is also done using Run-II data by the CMS collaboration, and it is observed that $m_{\tilde{g}} <$ 1.5 TeV is excluded \cite{CMS:2018ikp}. 
 
\item \textbf{One lepton and multi jets:}
Final states consisting of one lepton, different jet, and b jet multiplicities have been searched by the CMS \cite{cms_web1, cms_web2, CMS:2017szl} and the ATLAS \cite{ATLAS:2021fbt, ATLAS:2017oes, ATLAS:2021fbt} collaborations with Run-II data. Prompt decay of gluino via $\lambda^{\prime\prime}$ coupling has been considered and $m_{\tilde{g}} <$ 1.61 TeV have been excluded \cite{CMS:2017szl}. 
$m_{\tilde{g}}$ has been excluded up to 2.1 TeV and 1.8 TeV for different values of $m_{\lspone}$ corresponding to $\lambda^{\prime\prime}_{112}$ and $\lambda^{\prime}$ respectively by the ATLAS collaboration \cite{ATLAS:2017oes}. Similar searches have been conducted considering cascade decay of gluino, and it has been shown that $m_{\tilde{g}} < $ 1.36 TeV is excluded by the CMS collaboration \cite{cms_web2}. Similarly, in Ref.~\cite{ATLAS:2021fbt}, the limits on gluino masses are provided depending upon the possible decay modes of gluino in the presence of different non-zero couplings. $m_{\tilde{g}}$ is excluded upto 2.38 TeV when it decays into $\tilde{g} \rightarrow t \bar{t} \lspone \rightarrow t \bar{t}t b s$ with bino type LSP. Also, for the decay mode $\tilde{g} \rightarrow \bar{t} \tilde{t} \rightarrow \bar{t} b s$ the $m_{\tilde{g}}$ is excluded upto 1.83 TeV. Similarly, $m_{\tilde{g}}$ is excluded upto 2.25 TeV for the decay mode $\tilde{g} \rightarrow q \bar{q} \lspone \rightarrow q \bar{q} q \bar{q} l/\nu$. 
\end{itemize}
%%%%%%%%%
There are exclusion limits on gluino mass derived from the Run-I data of the LHC \cite{ATLAS:2014kpx, ATLAS:2014pjz, CMS:2016zgb, ATLAS:2015xmt, ATLAS:2015gky, ATLAS:2013qzt}, which are understandably weaker.
%%%%%%%%%%%%%%%%%%%%%%%%%%%%%%%%%%
\subsection{Squark search status}
%%%%%%%%%%%%%%%%%%%%%%%%%%%%%%%%%%
Now, we proceed to discuss the searches for squarks pair production. Similar to the gluino case, we categorize the searches depending on different possible final states and summarize the exclusion limits on squark masses.  
\begin{itemize}

\item \textbf{Two leptons and jets :} 
The final state consisting of two same-sign leptons, two jets, at least one of which is a $b$-jet has been searched by the experimental collaborations \cite{ATLAS:2017tmw, ATLAS:2023afl, CMS:2020cpy}. From this final state search, $m_{\tilde{d}_R}$ is excluded upto $\sim$ 500 GeV \cite{ATLAS:2017tmw} and the lightest stop quark is excluded upto 1.7 TeV~\cite{ATLAS:2023afl}. From a similar search done by the CMS collaboration, the lightest top and bottom squark are excluded up to 900 GeV \cite{CMS:2020cpy}. 

%The oppositely charged two leptons plus 2 $b$ jets final state search is also done by ATLAS~\cite{ATLAS:2017jvy} and \tcr{it is found the limit on $m_{\tilde{t}}$ between 600 GeV for large $b\tau$ decay branching ratios and 1500 GeV for a $be$ branching ratio of 100\%.} 

\item \textbf{Multi jets :}
The LHC collaborations have also considered the scenario where the top squark 
decays into quarks via the LSP \cite{ATLAS:2016lmi, ATLAS:2017jnp, ATLAS:2020wgq, CMS:2018mts, CMS:2018pdq}. This kind of scenario can arise when the 
$\lambda^{\prime\prime}$ couplings are non-zero. The final state consists of no leptons and at least four jets. The most stringent limit is obtained when the LSP is higgsino type. $m_{\tilde{t}}$ is excluded upto 950 GeV depending upon the values of $m_{{\tilde{\chi}}_{1,2}^0}$/$m_{{\tilde{\chi}}_1^{\pm}}$ \cite{ATLAS:2020wgq}. Top squark decaying into a bottom quark and a light quark leads to the exclusion limit $100 \leq m_{\tilde{t}} \leq 470$ and $480 \leq m_{\tilde{t}} \leq 670$ \cite{ATLAS:2017jnp}. Instead, when the top squark decays into two light quarks, the limit is modified to $100 \leq m_{\tilde{t}} \leq 410$ \cite{ATLAS:2017jnp}.

\item \textbf{One lepton and either zero or at least three $b$ jets :} 
The final state with at least one isolated lepton and either zero or three $b-$ tagged jets is searched by ATLAS collaboration \cite{ATLAS:2017oes, ATLAS:2021fbt}. Top squark mass has been excluded up to 1.12 TeV, 1.22 TeV, and 1.36 TeV for wino type LSP, higgsino type LSP, and bino type LSP respectively as shown in Fig.~\ref{fig:gluino_reach} \cite{ATLAS:2021fbt}.

\item \textbf{Two top quark and light jets :} 
The final state consists of two top quarks and several light jets. Here, the top squark is decaying to $\lspone$ and one top quark and $\lspone$ decays via $\lambda_{ijk}^{\prime\prime}$ coupling. From this search, $m_{\tilde{t}} <$ 670 has been excluded by the CMS collaboration \cite{CMS:2021knz}. 

\end{itemize}
%%%%%%%%%
There are exclusion limits on squark masses derived from the Run-I data of the LHC \cite{ CMS:2013pkf, ATLAS:2014kpx, CMS:2014tla, CMS:2014wpz, ATLAS:2015gky, ATLAS:2016lmi, CMS:2016ooq, cms_web3}, which are understandably weaker.

%%%%%%%%%%%%%%%%%%%%%%%%%%%%%%%%%%%%%%%%%%%%%%%%%%%
\subsection{Electroweakino search status}
\label{sec:electroweakino} 
%%%%%%%%%%%%%%%%%%%%%%%%%%%%%%%%%%%%%%%%%%%%%%%%%%
The electroweakino search is divided into two parts depending on the production modes. We first discuss the existing bounds on the neutralino and chargino masses before proceeding to sleptons and sneutrinos.
%%%%%%%%%%%%%%%%%%%%%%%%%%%%%%%%%%%%%%%%%%  
\subsubsection{Neutralino-Chargino search}
%%%%%%%%%%%%%%%%%%%%%%%%%%%%%%%%%%%%%%%%%%
Depending on the nature of the LSP and NLSP, the different probable final states originating from chargino-neutralino production are discussed in the Refs.~\cite{Dumitru:2018nct, Dumitru:2019cgf}. Similar to squark and gluino searches, here we also categorize the results according to the final states.

\begin{itemize}
 
\item \textbf{Lepton and multi jets :} 
The ATLAS collaboration has categorized signal region into two final states: one consisting of one lepton and six jets along with at least four b-jets, the other consisting of two same-sign leptons and six jets along with at least three b-jets \cite{ATLAS:2021fbt}. From this analysis, they have concluded that for the direct production of electroweakino ($\lspone\lsptwo$ and $\lspone\chonepm$), the wino (higgsino) masses are excluded between 197-365 (200-320) GeV~\cite{ATLAS:2021fbt}. 

\item \textbf{Four or more lepton search :} 
In this scenario, the LSP decays via non-zero $\lambda$ couplings to give rise to a large multiplicity of leptons in the final state.
There exists a few articles which explore similar final states \cite{ATLAS:2021yyr, ATLAS:2018rns, ATLAS:2014pjz}. The most stringent limit from this signal region is provided by the ATLAS collaboration \cite{ATLAS:2021yyr}. For wino like NLSP, $m_{\chonepm} = m_{\lsptwo}$ is excluded upto 1.6 TeV for $m_{\lspone} < $ 800 GeV corresponding to $\lambda_{12k}$ coupling. On the other hand when $\lambda_{i33}$ couplings are non-zero, $m_{\chonepm} = m_{\lsptwo}$ is excluded upto 1.13 TeV for 500$ < m_{\lspone} < 700$ GeV \cite{ATLAS:2021yyr}. This kind of search has also been performed by the CMS collaboration using Run-I data. They observe that the
 wino (higgsino) like neutralino ($\lspone$) is excluded between 700-875 (300-900) GeV \cite{CMS:2016zgb, CMS:2013pkf}.
 
\end{itemize}
%%%%%%%%%%%%%%%%%%%%%%%%%%%%%%%
\subsubsection{Slepton search}
%%%%%%%%%%%%%%%%%%%%%%%%%%%%%%%
A multilepton final state happens to be the most promising signal to look for sleptons and sneutrinos at the LHC. ATLAS collaboration has presented their results derived from both RUN-I and RUN-II data for signal regions with four or more leptons in the final state \cite{ATLAS:2014pjz, ATLAS:2018rns, ATLAS:2021yyr}. They have considered pair and associated production of mass degenerate left-handed sleptons and sneutrinos of all three generations ($\tilde{l}_L \tilde{l}_L$, $\tilde{\nu}_L \tilde{\nu}_L$, $\tilde{l}_L \tilde{\nu}_L$).   The possibility that the LSP ($\lspone$) arising from the cascades can eventually decay into charged leptons and neutrinos through non-zero $\lambda_{12k}$ and $\lambda_{i33}$ couplings has been explored in \cite{ATLAS:2014pjz, ATLAS:2018rns, ATLAS:2021yyr}. In the presence of non-zero $\lambda_{12k}$, they have excluded left-handed slepton or sneutrino masses up to 1.2 TeV for $m_{\lspone} \sim$ 800 GeV with RUN-II data \cite{ATLAS:2021yyr}. In the presence of non-zero $\lambda_{i33}$ couplings, the same sparticle masses are excluded up to 0.87 TeV for $m_{\lspone} \sim$ 500 GeV \cite{ATLAS:2021yyr}. 

%%%%%%%%%%%%%%%%%%%%%%%%%%%%%%%%
\subsection{Bilinear RPV model search status}
\label{sec:bilinear}
%%%%%%%%%%%%%%%%%%%%%%%%%%%%%%%%
Bilinear RPV couplings can give rise to a wide range of final states at the LHC. When the RPV parameters are large enough, we obtain substantial mixing among the neutralino-neutrino and chargino-charged lepton states. In the scalar sector, the charged sleptons mix with charged scalars and neutral scalars mix with sneutrinos. All these mixing terms give rise to some exotic signals that are novel and can be easily distinguished from RPC signatures \cite{deCampos:2007bn, deCampos:2012pf, Roy:1996bua, Hirsch:2003fe}. Here, we summarise LHC searches performed in this context. 

%%%%%%%%%%%%%%%%%%%%%%%%%%%%%%%%%%%%%%%%
\subsubsection{mSUGRA/CMSSM bRPV Model} 
%%%%%%%%%%%%%%%%%%%%%%%%%%%%%%%%%%%%%%%%
Three main signal regions are searched for in this context. From RUN-I data, the ATLAS collaboration has excluded 200 $< m_{1/2} <$ 500 GeV for $m_0 <$ 2.2 TeV looking at a final state with at least one lepton, many jets and missing transverse energy originating from squark and gluino production \cite{ATLAS:2015rul}. From similar production channels by looking at a final state consisting of two same-sign charged leptons or three leptons and jets, they have excluded 200 $< m_{1/2} < $490 GeV for $m_0 <$ 2.1 TeV \cite{ATLAS:2014kpx}. Another final state consisting of at least one $\tau$, multi-jets, and missing transverse energy arising from colored sparticle productions have excluded $m_{1/2}$ up to 680 GeV and 500 GeV for light $m_0 <$ 750 GeV and heavy $m_0 >$ 680 GeV respectively \cite{ATLAS:2014eel, ATLAS:2015gky}.   
%%%%%%%%%%%%%%%%%%%%%%%%%%%%%%%%%%  
\subsubsection{pMSSM bRPV Model} 
%%%%%%%%%%%%%%%%%%%%%%%%%%%%%%%%%%
The low-scale bilinear RPV scenario has been explored similarly through colored sparicles and electroweakino searches. Starting from top squark production, the ATLAS collaboration has excluded the $\mu$ parameter in the range 160 $< \mu <$ 455 GeV for light top squark mass $m_{\tilde{t}_1}$ around 500-580 GeV \cite{atlas_web2} assuming that after production it decays via higgsino-like neutralinos and chargino. Taking light squark production into account, following similar analysis, they have excluded $\mu <$ 560 GeV for $m_{\tilde{q}_{l,3}}$ = 800 GeV \cite{atlas_web2}. In the electroweak sector, $\chonepm \chonemp$ and $\chonepm \lspone$ productions and their subsequent decays $\chonepm \rightarrow l^{\pm} + Z$, $Z \rightarrow l^{\pm} + l^{\mp}$; $\chonemp \rightarrow Z/h/W^{\pm} + l^{\mp}/l^{\mp}/\nu$; $\lspone \rightarrow W^{\pm}/Z/h + l^{\mp}/\nu/\nu$ have been considered. They have defined three different signal regions depending upon the number of leptons and the presence of leptonically or hadronically decaying Z bosons. They have presented the exclusion limit on masses of higgsino type $\chonepm$/$\lspone$ depending on their branching ratio to Z boson \cite{ATLAS:2020uer}. The exclusion limits on the NLSP masses are at 625 GeV, 1050 GeV, and 1100 GeV for 100\% branching ratio to $Z+\tau$, $Z+\mu$, and $Z+e$ channel respectively \cite{ATLAS:2020uer}. In Ref.~\cite{ATLAS:2023lfr} they have considered $\chonepm\lspone + \lspone\lsptwo$ production followed by decays $\chonepm \rightarrow W^{\pm} + \nu$ and $\lspone$/$\lsptwo \rightarrow W^{\pm} + l^{\mp}$. They have explored the signal region with two same-sign leptons or three leptons. From this search, $\lspone$/$\lsptwo$/$\chonepm <$ 440 GeV has been excluded.

\subsection{Search for long-lived particle}
Most of the LHC analyses that we have already discussed are based on the 
assumption that the SUSY particles decay promptly within the detector. In recent 
years, searches for long-lived particles (LLP) have gained more attention from the community. The LLPs are very common in RPV scenarios when either the RPV couplings or the mass gap between NLSP and LSP is very small. The LLPs 
produce various unconventional detector signatures and in many scenarios, they 
are often free from the SM irreducible backgrounds. For example, the LLP pair production may give rise to two displaced vertices which are formed from 
the intersection of various charged tracks. These vertices are displaced from the 
beam axis but identified within the radius of the beam pipe. Apart from displaced 
vertices, the ATLAS and CMS collaborations have also used signatures like displaced jets/leptons, disappearing tracks, non-pointing photons, etc. It may be noted that 
the derived exclusion limits crucially depend on the mean proper lifetime or the proper decay length.

In a recent analysis ~\cite{ATLAS:2023oti}, the ATLAS Collaboration has looked for the direct pair production of pure higgsino-like electroweakinos and pair production of gluinos where each gluino decays promptly into a neutralino (LLP) and $q \bar{q}$ pair with 100\% branching ratios. These electroweakinos further decay to light flavor quarks via the UDD type RPV couplings and generate the signature of displaced vertices. Electroweakinos with masses below 1.5 TeV are excluded for mean proper lifetime between 0.03 ns ($c\tau$ = 0.9 cm) to 1 ns ($c\tau$ = 30 cm)~\cite{ATLAS:2023oti}. From gluino pair production (with a fixed choice of gluino mass = 2.4 TeV), electroweakinos with masses below 1.5 TeV are excluded for mean proper lifetime between 0.02 ns to 4 ns~\cite{ATLAS:2023oti}. 
The limits on the displaced gluinos are more stringent. The CMS collaboration 
has excluded gluino mass up to 2.5 TeV~\cite{CMS:2020iwv}. From the stop pair production with LQD-type couplings, displaced stops are excluded below 
1.6 TeV~\cite{CMS:2020iwv}. In a recent phenomenological work~\cite{Bhattacherjee:2023kxw}, the searching prospect of long-lived LSP has been explored at the HL-LHC 
from electroweakino pair production. It has been observed that 
the wino-like ${\lsptwo/\chonepm}$ can be probed up to 1.9 TeV with $m_{\lspone} > 800$ across a decay length ranging from 1 cm to 
200 cm. 
For the higgsino like pair production, $m_{\lsptwo/\chonepm}$ can be probed 
upto 1.6 TeV for  $m_{\lspone} > 700$ across a decay length range 1-200 $cm$
~\cite{Bhattacherjee:2023kxw}.

%%%%%%%%%%%%%%%%%%%%%%%%%%%%%%%%%%%%%%%%%%%%%%%%%%%
\subsection{Resonance Search}
\label{sec:res}
\begin{figure}[!htb]
\begin{center}
\includegraphics[width=0.4\textwidth]{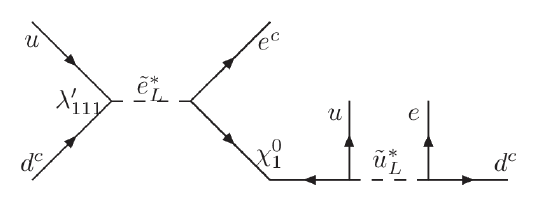} \hspace{2cm}
\includegraphics[width=0.4\textwidth]{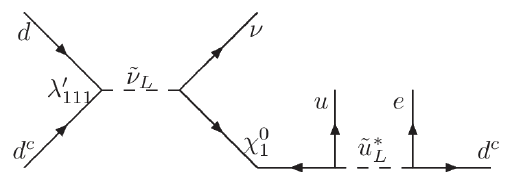}
\caption {Representative diagrams contributing to resonance search of sleptons and sneutrinos in the presence of non-zero $\lambda^{\prime}_{111}$ coupling.}
\label{fig:res}
\end{center}
\end{figure}

\begin{figure}[!htb]
\begin{center}
\includegraphics[width=7cm,height=6cm]{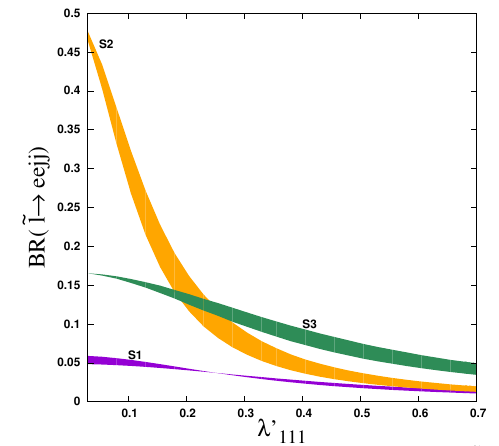} 
\includegraphics[width=7cm,height=6cm]{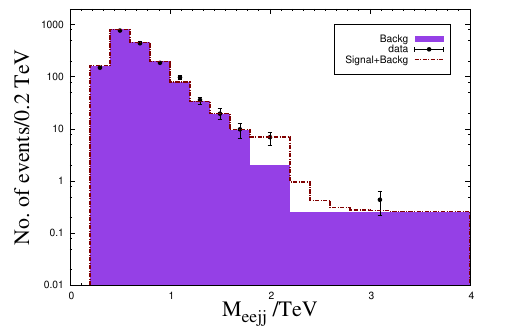}
\caption {Effective branching ratio ${\rm BR}(\tilde{l}\to eejj)$ for the three scenarios considered with varied $m_{\tilde\chi_1^0}=532$ (400-1000 GeV) and fixed $\lambda^{\prime}_{111}$ (left) and $M_{eejj}$ distribution after imposing all cuts on signal and background events corresponding to {\bf S3} scenario with $\lambda^{\prime}_{111}=0.105$ and $m_{\tilde\chi_1^0}=532$ GeV (right) \cite{Allanach:2014lca}.}
\label{fig:res1_eejj}
\end{center}
\end{figure}

%%%%%%%%%%%%%%%%%%%%%%%%%%%%%%%%%%%%%%%%%%%%%%%%%%%
Supersymmetric particles can be produced through RPV couplings at the LHC. Sleptons and sneutrinos single production through non-zero $\lambda^{\prime}$ couplings have been studied by the ATLAS and CMS collaborations \cite{CMS:2018skt,ATLAS:2012llm,ATLAS:2015dva}. After being produced, the sleptons or the sneutrinos can decay via $\lambda$ couplings, resulting in some very distinguishable same-sign lepton signatures \cite{CMS:2018skt,ATLAS:2012llm,ATLAS:2015dva}. One can also have various non-zero $\lambda^{\prime}$ couplings appearing both at the production and decay vertices of the sleptons and sneutrinos, which can mimic final states arising from heavy gauge boson \cite{CMS:2014nrz} or a di-leptoquark production \cite{CMS:2014qpa}. Subsequent decays of these particles can produce a one(two) lepton(s) two jets final state. Such possibilities have been studied in the context of non-zero $\lambda^{\prime}_{111}$ coupling \cite{Allanach:2014lca,Allanach:2014nna}. The representative Feynman diagrams are shown in Fig.~\ref{fig:res}. Similar final states can arise with charginos in the cascade as well.  
%%%%%%%%%%%%%%%%%%%%%%%%%%%%%%%%%%%%%%%%%%%%%%%%%%%%%%%
%%%%%%%%%%%%%%%%%%%%%%%%%%%%%%%%%%%%%%%%%%%%%%%%%%%%%%%
Depending on the nature of the lightest neutralino or chargino, the decay branching ratios of the SUSY particles change. Three separate scenarios were considered: $M_1 < M_2 = M_1 + 200 < \mu$ ({\bf S1}), $M_1 < \mu < M_2$ ({\bf S2}) and $M_2 << M_1\simeq \mu$ ({\bf S3}). The effective branching ratio ${\rm BR}(\tilde{l}\to eejj)$ is shown on the left in Fig.~\ref{fig:res1_eejj}. The color bands indicate the small variation of the effective BR with the lightest neutralino mass varied within the window 400 - 1000 GeV for a fixed $\lambda^{\prime}_{111}$.

The final state consists of exactly two isolated leptons and at least two jets. Alongside the basic selection criteria, the final states were selected with an invariant mass on the lepton pair $M_{ee} > 200$ GeV and on the leptons and two hardest jets $M_{eejj} > 600$ GeV. Fig.~\ref{fig:res1_eejj} shows the yields per 200 GeV bins of $M_{eejj}$ on the right. The CMS collaboration reported a small excess around $1.8 {~\rm TeV} < M_{eejj} < 2.2 {~\rm TeV}$ \cite{CMS:2014nrz} which could be nicely explained within the RPV scenario. The results shown in this figure corresponds to the scenario {\bf S3} with $\lambda^{\prime}_{111}=0.105$ and $m_{\tilde\chi_1^0}=532$ GeV \cite{Allanach:2014lca}. 
%%%%%%%%%%%%%%%%%%%%%%%%%%%%%%%%%%%%%%%%%%%%%%%%%%%%%%%
%%%%%%%%%%%%%%%%%%%%%%%%%%%%%%%%%%%%%%%%%%%%%%%%%%%%%%%

\begin{figure}[!htb]
\begin{center}
\includegraphics[width=7cm,height=6cm]{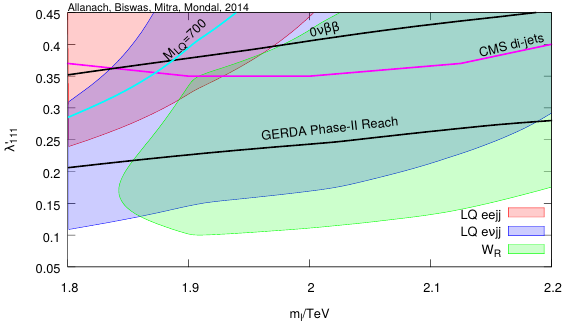}
\includegraphics[width=7cm,height=6cm]{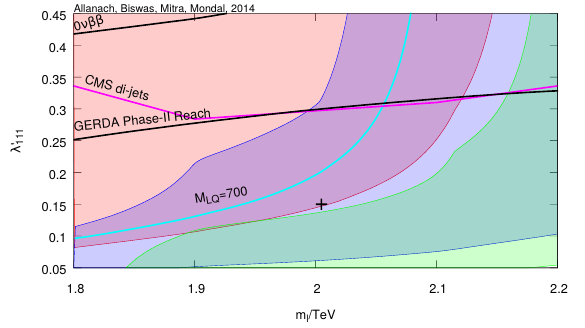}
\caption {Constrained parameter space in $m_{\tilde{l}}$ - $\lambda^{\prime}_{111}$ plane assuming $m_{\tilde{\chi}_1^0} = 900$ GeV for {\bf S1} and {\bf S2} scenario. The same color coding has been followed in both figures \cite{Allanach:2014nna}.}
\label{fig:fit_ejS1S2}
\end{center}
\end{figure}

In \cite{Allanach:2014nna}, the excess obtained in $eejj$ and $e\nu jj$ final state from leptoquark search as well as the $eejj$ excess obtained in heavy gauge boson search were explained together within RPV scenario with non-zero $\lambda^{\prime}_{111}$. Stringent constraint on the $\lambda^{\prime}_{111}$ coupling arises from neutrinoless double beta decay ($0\nu\beta\beta$), which was also taken into account to determine the available parameter space. For $0\nu\beta\beta$ both existing limit \cite{GERDA:2013vls} as well as projected limit from GERDA Phase-II \cite{Smolnikov:2008fu} were considered. CMS dijet search results \cite{CMS:2013egk} were also considered in order to constrain the parameter space. Compatible parameter space was found only for the {\bf S1} and {\bf S2} scenarios, as shown in Fig.~\ref{fig:fit_ejS1S2}. It is ensured that all the existing constraints are satisfied in the compatible parameter space while the excesses fit within their $95\%$ confidence level intervals. 
%%%%%%%%%%%%%%%%%%%%%%%%%%%%%%%%%%%%%%%%%%%%%%%%%%%%%%%
%%%%%%%%%%%%%%%%%%%%%%%%%%%%%%%%%%%%%%%%%%%%%%%%%%%%%%%
As evident, a very small parameter space was obtained for {\bf S1} that could satisfy all excesses subjected to other relevant constraints. The compatible parameter space (around $\lambda^{\prime}_{111}\sim 0.32$ and $m_{\tilde{l}}\sim 1.88$ TeV) would be further probed by $0\nu\beta\beta$ experiments. Marginal compatible parameter space was obtained for {\bf S2} as well. Part of the overlap regions (around $0.11 < \lambda^{\prime}_{111} < 0.13$ and $1.9 {~\rm TeV} < m_{\tilde{l}} < 2$ TeV) that would be probed by GERDA Phase-II is already disfavored by the CMS di-jet searches. 

%%%%%%%%%%%%%%%%%%%%%%%%%%%%%%%%%%%%%%%%%%%%%%%%%%%%%%
\subsection{Prospect of electroweakino searches at the HL-LHC and HE-LHC}
\label{sec:electroweakino_pheno}
%%%%%%%%%%%%%%%%%%%%%%%%%%%%%%%%%%%%%%%%%%%%%%%%%%%%
As the LHC Collaborations have already started collecting Run-III data, it is 
very crucial to gauge the reach of the SUSY parameter space which can be 
probed at the highest possible luminosity (HL-LHC) or the proposed high 
energy upgrade of the LHC (HE-LHC). The electroweak SUSY sector with the light 
charginos, sleptons, and neutralinos are highly motivated in the context of 
recent measurements of muon (g-2) \cite{Muong-2:2023cdq}. In Sec.\ref{sec:mug2} 
we have already discussed how RPV scenarios give rise to additional contributions 
apart from MSSM  contributions (neutralino-smuon and chargino-sneutrino loops). 
Depending upon the non-zero couplings and the LSP-NLSP mass difference, there can 
be a plethora of final states coming from the direct productions of 
electroweakinos.  In this section, we summarize a few recent works 
\cite{Choudhury:2023eje, Choudhury:2023yfg, Barman:2020azo} which have explored 
the prospect of electroweakino searches at the HL-LHC and HE-LHC. 
The works in Ref.~\cite{Choudhury:2023eje,Choudhury:2023yfg} have considered 
the  $LLE$ type couplings while $UDD$ type coupling has been studied  in 
\cite{Barman:2020azo}.

 \begin{figure}[h]
\begin{center}
\includegraphics[width=0.33\linewidth]{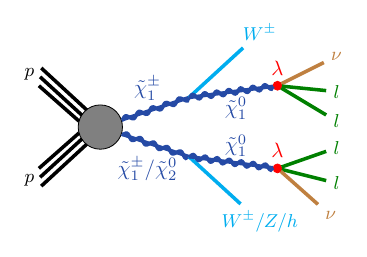}
\includegraphics[width=0.3\linewidth]{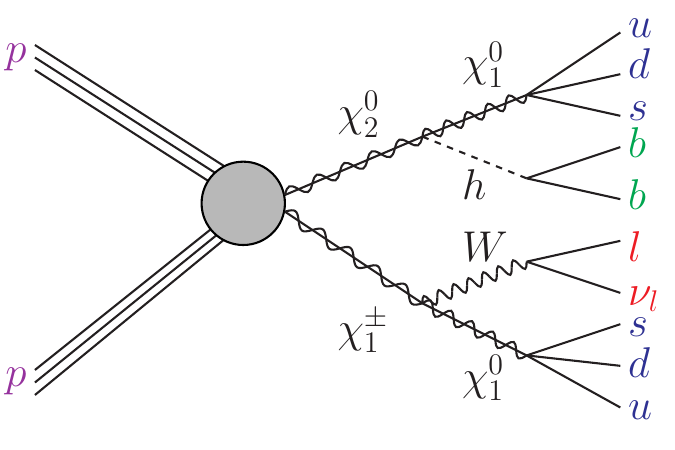}
\includegraphics[width=0.33\linewidth]{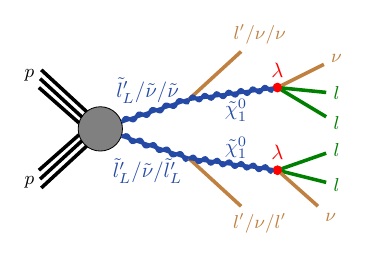}
\caption{Diagrams of wino pair production and decays of $\lspone$  via LLE coupling (left) \cite{Choudhury:2023eje} and UDD coupling (middle) \cite{Barman:2020azo}.  In the right panel slepton pair production and consequent decays of the LSP via LLE couplings are presented \cite{Choudhury:2023yfg}.  Here $l^{\prime} = e, \mu , \tau$ and $l = e, \mu$}
\label{fig:rpv_decay}
\end{center}
\end{figure}

There are 9 independent non-zero LLE type $\lambda_{ijk}$ couplings and 
depending upon the choices for a single non-zero $\lambda_{ijk}$ coupling there 
could be different lepton configuration arising from the LSP pair (see Fig.\ref{fig:rpv_decay}) which originates from the electroweakino pair production. The collider limits are not sensitive to flavors of the leptons as long as only electrons and muons are present in the final state \cite{ATLAS:2014pjz}. Considering leptons only as $l = e , \mu$, these non zero 
$\lambda_{ijk}$ couplings can give rise to four different scenarios. Refs.~\cite{Choudhury:2023eje, Choudhury:2023yfg} have considered all these four scenarios 
and derived the discovery reaches and the projected exclusion limits on $m_{\chonepm}$ and $m_{\tilde{l}^{\prime}_L}$/$m_{\tilde{\nu}_L}$ ($l^{\prime} \equiv e, \mu, \tau$) plane
using $N_l \geq$ 4 ($l\equiv e,~\mu$) final state.  
Among these four scenarios, for nonzero $\lambda_{121}$ and/or $\lambda_{122}$, 
the LSP pair always gives $4l$ final state with 100\% branching ratio and additional leptons 
may come from the $W/Z$ boson or slepton decay. It is also expected that  
scenarios with non zero $\lambda_{121}$ and/or $\lambda_{122}$ couplings will 
give the best possible projected limit and in this review, we only focus 
on this optimal scenario. 
 
%%%%%%%%%%%%%%%%%%%%%%%%%%%%%%%%%%%
\begin{figure}[!htb]
\begin{center}
\includegraphics[width=0.454\textwidth]{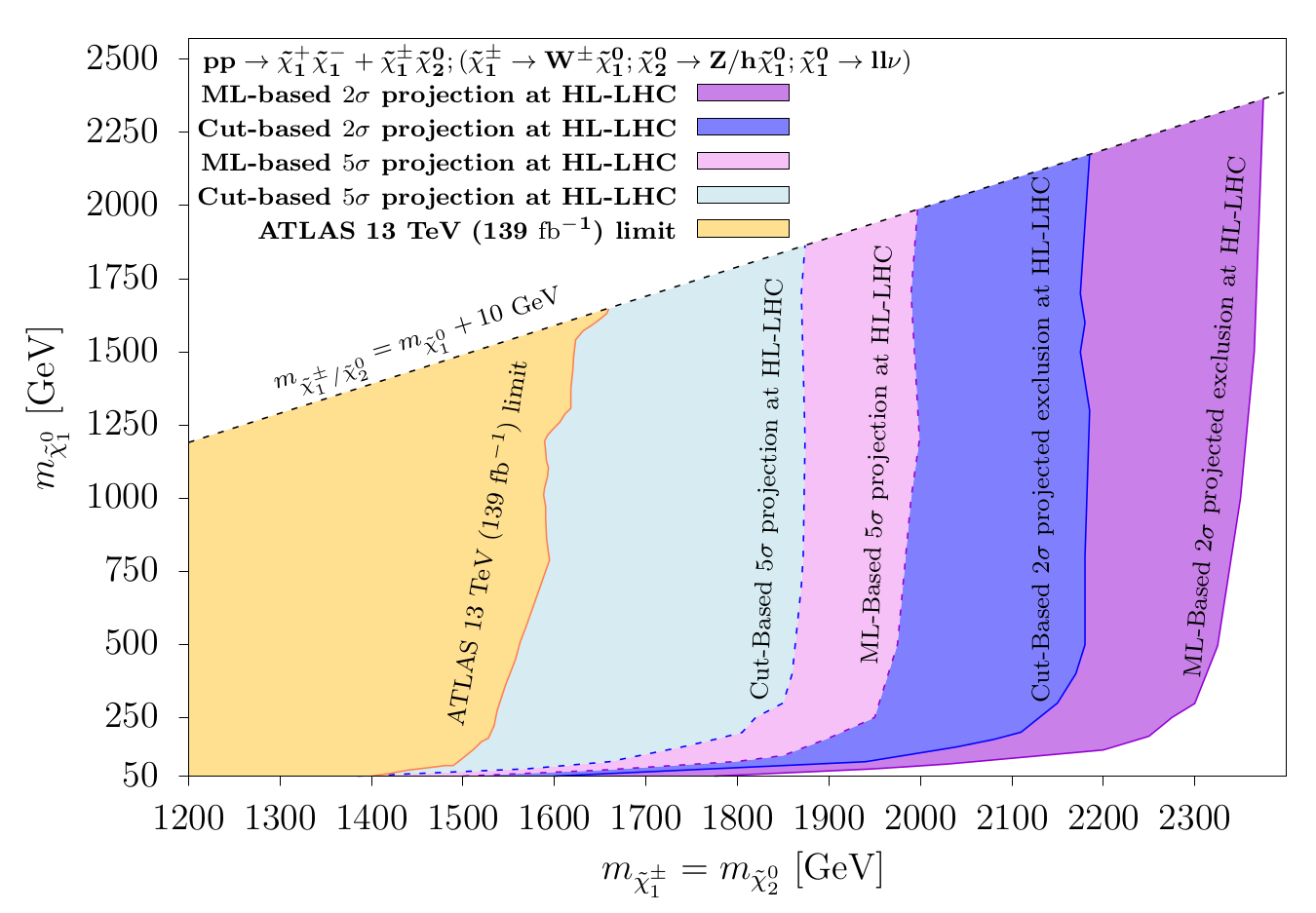}
\includegraphics[width=0.454\textwidth]{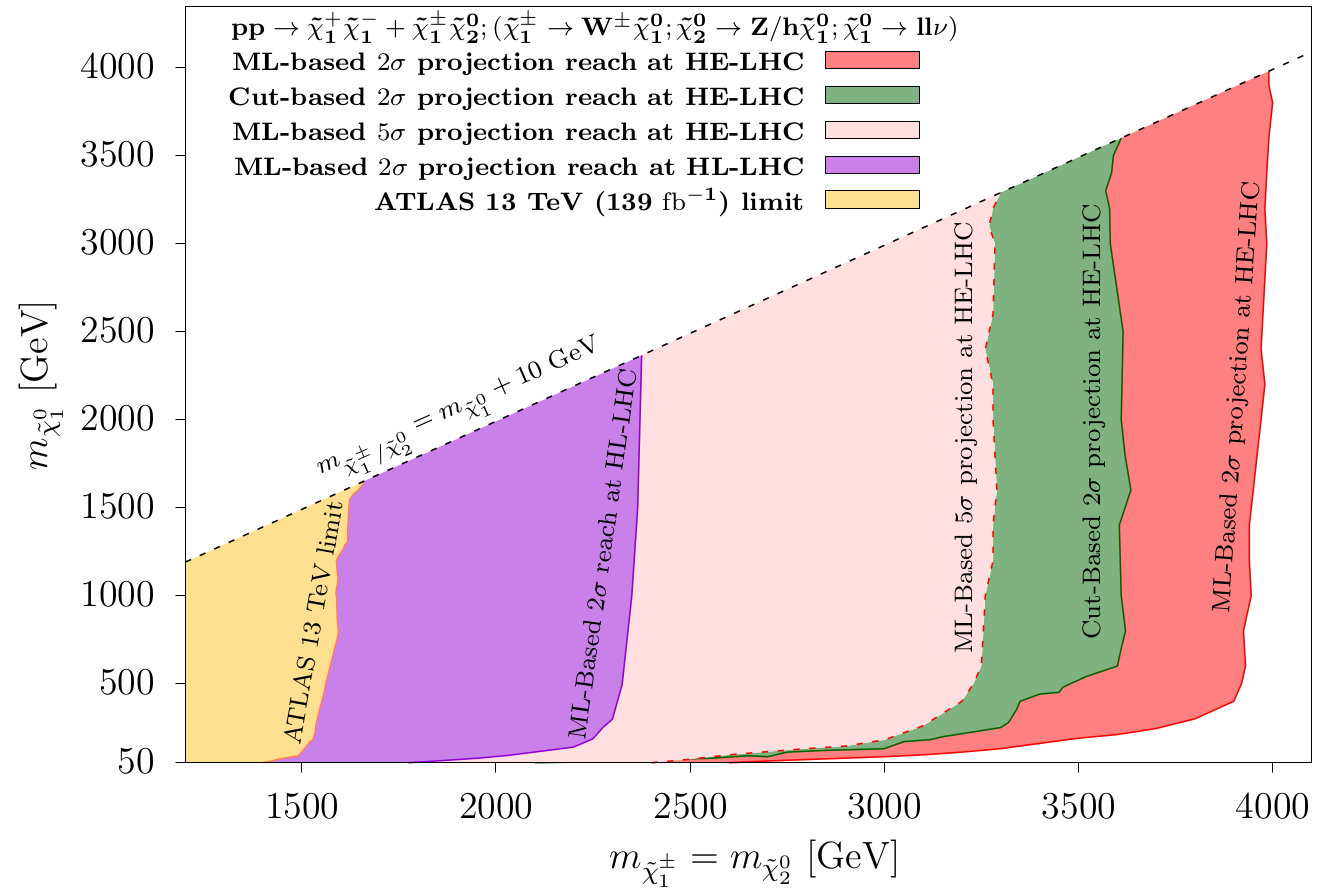}
\caption { Projected discovery ($5\sigma$) and exclusion ($2\sigma$) reaches in the $m_{\lsptwo=\chonepm} - m_{\lspone}$ mass plane at the HL-LHC (left) and at the 
HE-LHC (right) (from Ref.\cite{Choudhury:2023eje}). The light (dark) blue color refers to the $5\sigma$ ($2\sigma$) reach 
obtained from cut-based analysis. The light (dark) violet colors represent the $5\sigma$ ($2\sigma$) reach derived from the ML-based analysis at the HL-LHC.  
The light green color in the right panel represents the 2$\sigma$ projection coming from cut-based analysis for the HE-LHC.  The light (dark) red color corresponds to the $5\sigma$ ($2\sigma$) reach at the HE-LHC provided by the ML-based analysis. The yellow regions for both the figures represent the existing limits obtained by the ATLAS collaboration from Run-II data~\cite{ATLAS:2021yyr}.}
\label{fig:electroweakino_reach}
\end{center}
\end{figure}
%%%%%%%%%%%%%%%%%%%%%%%%%%%%%%%%%%%

Two production channels $pp\to \chonepm \lsptwo$ and $pp\to \chonep \chonem$ 
are considered in Ref.~\cite{Choudhury:2023eje} where the NLSP and the LSP 
are assumed to be wino-like and bino-like respectively (see Fig.\ref{fig:rpv_decay}
(left)). The projected discovery and exclusion reaches 
of the HL-LHC ($\sqrt{s} = 13$ TeV and $\mathcal{L} = 3000$ fb$^{-1}$) and HE-LHC ($\sqrt{s} = 27$ TeV and $\mathcal{L} = 3000$ fb$^{-1}$) 
obtained in~\cite{Choudhury:2023eje} are illustrated in the left and right panel of Fig.\ref{fig:electroweakino_reach} respectively. It is shown that a 
machine learning (ML) based multivariate analysis using an Extreme Gradient Boosted decision tree algorithm further improves the results compared to the traditional cut-based analysis. The projected $2\sigma$ exclusion limit reaches up to 2.37 TeV and 4.0 TeV at the HL-LHC and HE-LHC respectively from the ML-based analysis. The ML-based analysis provides an improvement of  $\sim$ 190 GeV (380 GeV) in the projected $2\sigma$ limits on the NLSP masses compared to the conventional cut-and-count analysis at the HL-LHC (HE-LHC).

\begin{figure}[!htb]
\begin{center}
\includegraphics[width=0.454\textwidth]{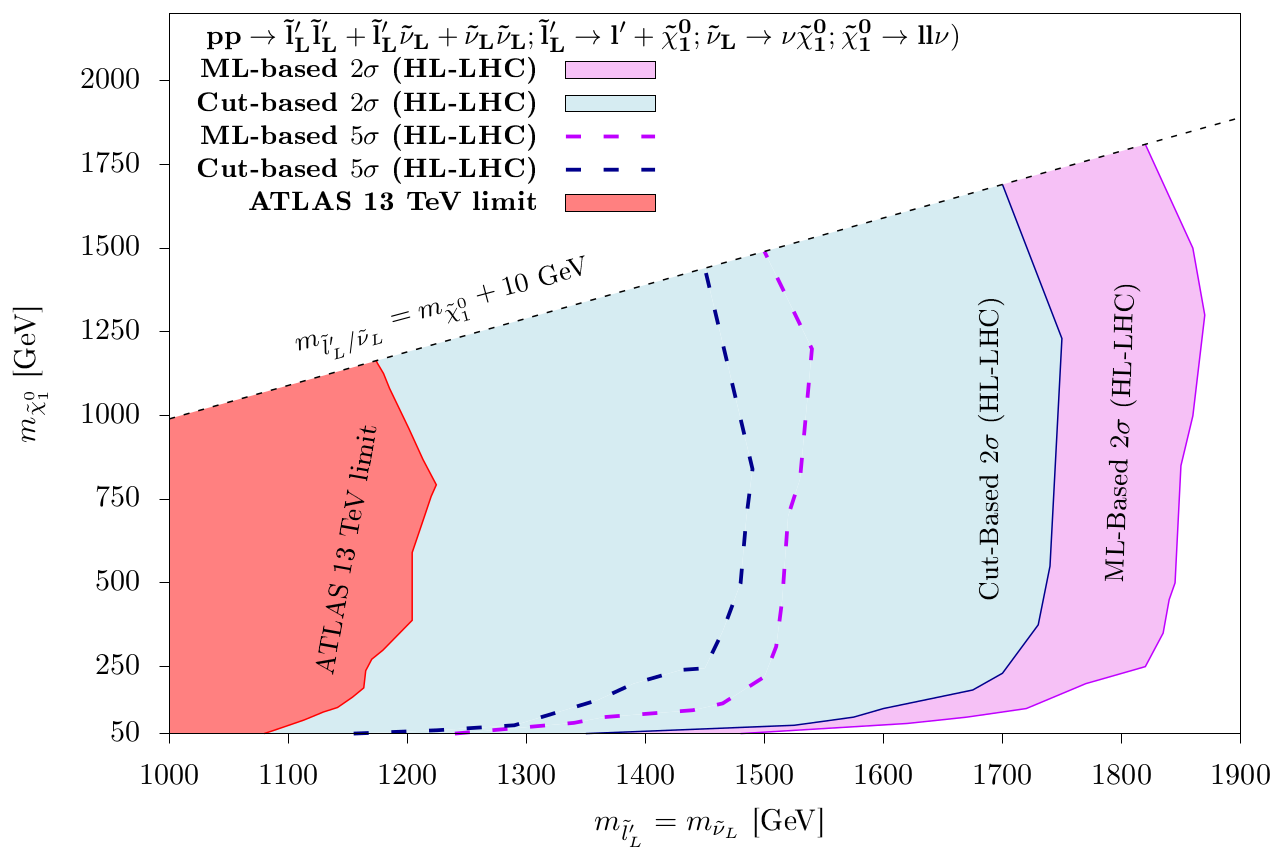}
\includegraphics[width=0.454\textwidth]{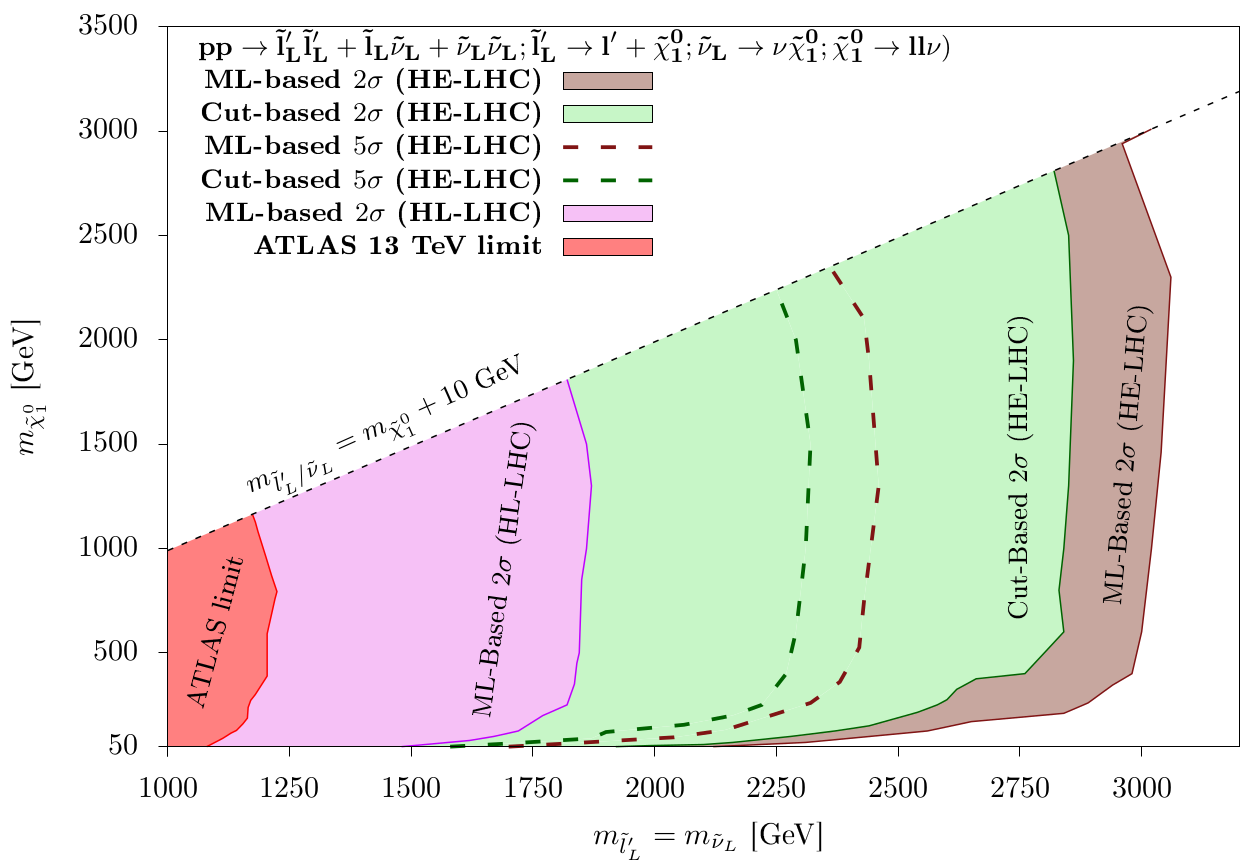}
\caption {Projected $5\sigma$ discovery and $2\sigma$ exclusion reach in the  $m_{\tilde{l}^{\prime}_L}-\mlspone$  mass plane at the  HL-LHC (Left) and 
 HE-LHC (Right) are presented here. The red regions for both the figures represent the existing limits obtained by the ATLAS collaboration from Run-II data\cite{ATLAS:2021yyr}. 
 The light blue and light violet colors represent the projected exclusion limits 
 using cut-based and ML-based analysis respectively at the HL-LHC. The light green and light brown colors refer to the $2\sigma$ regions provided from cut-based and ML-based analysis respectively at the HE-LHC. The dashed line with a similar color corresponds 
 to $5\sigma$ reach.}
\label{fig:slepton_reach}
\end{center}
\end{figure}

In another recent work  ~\cite{Choudhury:2023yfg}, the sensitivity of $N_l \geq$ 4 ($l\equiv e,~\mu$) final state has been studied for slepton pair productions at the HL-LHC 
and HE-LHC. In this work, both pair production and associated production of the three generations of left-handed charged sleptons and sneutrinos, which are assumed to be mass degenerate, have been considered. The diagram of the slepton production and consequent decays is shown in the right panel of Fig.\ref{fig:rpv_decay}. The 
discovery and exclusion reach on L-type slepton masses obtained for 
nonzero $\lambda_{121}$ and/or $\lambda_{122}$ coupling values are presented 
in Fig.\ref{fig:slepton_reach}. It is observed that 
the projected exclusion limits on slepton and sneutrino masses at the HL-LHC 
 are $\sim$ 1.85 TeV from ML-based analysis and $\sim$ 1.73 TeV from cut-based analysis while the HE-LHC analysis predicts an exclusion limits 
 upto $\sim$ 2.75 TeV and $\sim$ 3.0 TeV from cut-based and ML-based analysis respectively. The improvement in mass limits via the ML algorithm is better in the HE-LHC case. 

%%%%%%%%%%%%%%%%%%%%%%%%%%%%%%%%%%%
%%%%%%%%%%%%%%%%%%%%%%%%%%%%%%%%%%%%%%%%%%%

\begin{figure}[!htb]
\begin{center}
\includegraphics[width=0.459\linewidth]{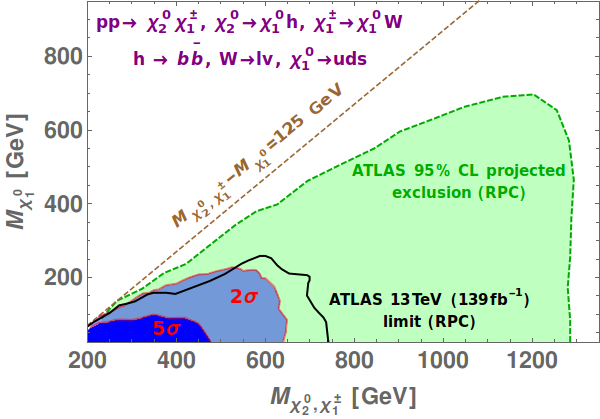}
\includegraphics[width=0.459\linewidth]{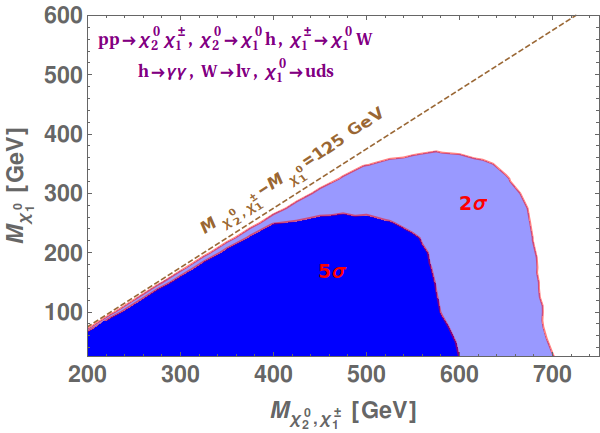}
\includegraphics[width=0.459\linewidth]{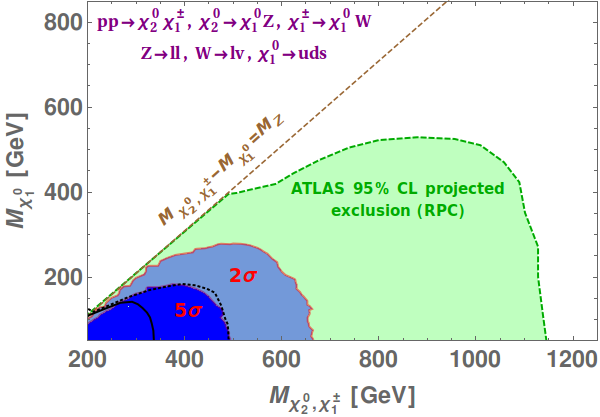}
\includegraphics[width=0.459\linewidth]{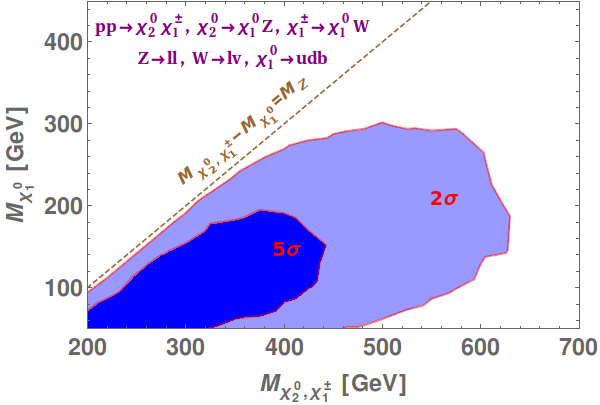}
\caption {The projected exclusion and discovery reaches (from \cite{Barman:2020azo}) are shown here with light and dark blue colors respectively 
using  the final states : $Wh$ mediated $1l + 2b + N_j \geq 2 + \met$ (Top left), 
 $Wh$ mediated $1l + 2\gamma + N_j \geq 2 + \met$ (Top right), 
 $WZ$ mediated $3l + N_j \geq 2 + \met$  (Botton left) 
 and $WZ$ mediated $3l + 2b + N_j \geq 2 + \met$ (Bottom right). The light green color corresponds to the 95\%$C.L.$ exclusion reach obtained at HL-LHC by the ATLAS collaboration considering the RPC SUSY framework \cite{ATLAS:2018diz}. }
\label{fig:udd_reach}
\end{center}
\end{figure}

The RPV SUSY scenarios with non-zero $\lambda_{ijk}$ couplings 
associated with LLE operators give rise to lepton-enriched final 
states from the LSP decays and  thus the limits on the electroweakinos masses 
are more stringent compared to the RPC SUSY scenarios. In the presence of 
baryon number violating UDD operators, the LSP decays to three quarks, and the 
jet-enriched final states are expected to provide a weaker limit compared to 
the conventional RPC scenarios. In Ref.~\cite{Barman:2020azo}, 
RPV scenarios with $\lambda_{112}^{\prime \prime}u^{c}d^{c}s^{c}$ 
and $\lambda_{113}^{\prime\prime}u^{c}d^{c}b^{c}$ interactions have been 
considered and the projected reach of direct electroweakino searches at the HL-LHC 
is presented considering four different final states. 
In this analysis \cite{Barman:2020azo}, $\chonepm\lsptwo$ pair production is considered where the wino like  $\chonepm$ and $\lsptwo$ decay as  
$\chonepm \to W \lspone$ and  $\lsptwo \to Z/h \lspone$ with 100\% branching ratios. The considered final states are as follows: 
\begin{itemize}
\item \textit{$Wh$ mediated $1l+2b+jets+\met$} final state coming from the decay chain  $\chonepm\lsptwo$ $\to$ $\left(W \lspone\right) 
\left(h \lspone\right)$ $ \to 
\left( l \nu uds \right)\left(b b u d s\right)$ (illustrated in Fig.\ref{fig:rpv_decay} (middle)). The corresponding projected 2$\sigma$ and 5$\sigma$ reach (from \cite{Barman:2020azo}) are presented in Fig.~\ref{fig:udd_reach} (top left).  
\item \textit{$Wh$ mediated $1l+2\gamma+jets+\met$}  
final state arising from the decay chain 
$\chonepm\lsptwo$ $\to$ $\left(W \lspone\right) 
\left(h \lspone\right)$ $ \to 
\left( l \nu uds \right)\left(\gamma \gamma u d s\right)$. The projected 2$\sigma$ and 5$\sigma$ reach for this channel obtained in \cite{Barman:2020azo}  are presented in Fig.~\ref{fig:udd_reach} (top right).
\item \textit{$WZ$ mediated $3l+jets+\met$ } 
final state emerges from the process  
$ \chonepm\lsptwo$ $\to$ $\left(W \lspone\right) 
\left(Z \lspone\right)$ $ \to 
\left( l \nu uds \right)\left(l l u d s\right)$. We present the 2$\sigma$ and 5$\sigma$ projection contour in Fig.~\ref{fig:udd_reach} (bottom left).
\item \textit{$WZ$ mediated $3l+2b+jets+\met$}  final state originating from the decay chain $ \chonepm\lsptwo$ $\to$ $\left(W \lspone\right)$ $\left(Z \lspone\right)$ $ \to \left(l \nu udb \right)$ $\left( ll u d b\right)$. 
The projected 2$\sigma$ and 5$\sigma$ reach for this scenario are presented in Fig.~\ref{fig:udd_reach} (bottom right).
\end{itemize}

It is observed that in the first three cases the projected exclusion contour reach up to $600-700~{\rm GeV}$ for a massless bino-like $\lspone$, while the last one with $\lambda_{113}^{\prime\prime}u^{c}d^{c}b^{c}$ operator
provides a projected exclusion reach up to $600~{\rm GeV}$ for $150~{\rm GeV} < M_{\lspone} < 250~{\rm GeV}$. The baryon number violating simplified scenarios  
considered in the work \cite{Barman:2020azo} are found to furnish a weaker projected reach (typically by a factor of $\sim 1/2$) than the 
RPC scenarios~\cite{ATLAS:2018diz}. 

%%%%%%%%%%%%%%%%%%%%%%%%%%%%%%%%%%%%%%%%%%
\section{Summary}
\label{sec:concl}
%%%%%%%%%%%%%%%%%%%%%%%%%%%%%%%%%%%%%%%%%%
Supersymmetry remains one of the most highly motivated BSM scenarios from the theoretical standpoint. Allowing R-parity violation within the theory extends the minimal version of SUSY, namely, RPC SUSY, by introducing four more interaction terms in the superpotential. This results in several novel phenomenological implications. In this article, we have discussed some of these phenomenological aspects in the context of light neutrino mass generation, muon anomalous magnetic moment, various lepton and quark flavor violating (QFV) decays, and collider searches. We have discussed how the RPV couplings affect light neutrino masses and mixings and thereby, can explain the neutrino oscillation data. We proceed to explore the case of the bilinear RPV scenario and showcase how strongly the experimental data from the neutrino and Higgs sector can constrain the RPV parameters. We have discussed possible RPV contributions to the lepton magnetic moment arising from the bilinear and lepton number violating trilinear couplings that can explain the observed anomaly in the measurement of muon (g-2). Taking into account the updated data of the measurement of muon (g-2) one can restrict products of trilinear RPV couplings very effectively. Possible RPV contributions to LFV and QFV decays and how they can restrict the values of the RPV couplings have been discussed next. We have summarized the LHC exclusion limits provided by the ATLAS and CMS collaborations in the context of RPV models. We have listed the most updated bounds on all SUSY particles arising from different final states. In this context, we highlight the prospect of resonance searches in the presence of non-zero $\lambda^{\prime}$ couplings. We have also discussed the future prospect of gaugino and slepton/sneutrino searches in different channels with varied lepton multiplicities in the presence of non-zero $\lambda$ and $\lambda^{\prime\prime}$ couplings in the context of HL-LHC and HE-LHC. 
These results are useful to understand how much of the vast RPV parameter space has already been probed by the existing data and at the same time provide an estimate of the available regions which should be probed further.   

  \vspace{+1cm}

 \noindent
\textbf{Acknowledgments:} The works described in Sec.~\ref{sec:neutrino}, 
Sec.~\ref{sec:res} and Sec.~\ref{sec:electroweakino_pheno} are based on 
Refs.~\cite{Choudhury:2023lbp, Choudhury:2023eje, Choudhury:2023yfg, Barman:2020azo, Allanach:2014lca,Allanach:2014nna}. We thank Ben Allanach, Rahool Kumar Barman,  Biplob Bhattacherjee, Sanjoy Biswas, Indrani Chakraborty, Manimala Mitra, Sourav Mitra, Najimuddin Khan and Subhadeep Sarkar for discussions and collaboration.

 \vspace{+1cm}

\noindent
\textbf{Data availability statement:} No data associated in the manuscript.

%%%%%%%%%%%%%%%%%%%%%%%%%%%%%%%%%%%%%%%%%% 
\bibliography{reference}

\end{document}